\documentclass[prd, aps, twocolumn, nofootinbib, showpacs, superscriptaddress]{revtex4-2}
\usepackage[colorlinks=true,citecolor=green,urlcolor=blue,linkcolor=blue]{hyperref}
\usepackage[utf8]{inputenc}
\usepackage{graphicx,mathtools,xcolor,latexsym,rotating,soul}
\usepackage[T1]{fontenc}
\usepackage{tikz}
\usepackage{amsmath, amssymb, amsthm, amsfonts,breqn}
\usetikzlibrary{decorations.pathmorphing}

\newtheoremstyle{named}{}{}{\itshape}{}{\bfseries}{.}{.5em}{\thmnote{#3 }#1}
\theoremstyle{named}

\newcommand*{\rom}[1]{\expandafter\@slowromancap\romannumeral #1@}

\newcommand\T{\rule{0pt}{2.5ex}}       
\newcommand\B{\rule[-2.5ex]{0pt}{0pt}}

\newcommand{\R}{\mathcal{R}}
\newcommand{\detP}{\text{det}\, \mathbb{P}}

\makeatletter
\let\cat@comma@active\@empty
\makeatother

\begin{document}
\title{Where and why does Einstein-Scalar-Gauss-Bonnet theory break down?}
\author{Abhishek Hegade K R}
\affiliation{Illinois Center for Advanced Studies of the Universe, Department of Physics, University of Illinois at Urbana-Champaign, Urbana, IL 61801, USA}

\author{Justin L. Ripley}
\affiliation{Illinois Center for Advanced Studies of the Universe, Department of Physics, University of Illinois at Urbana-Champaign, Urbana, IL 61801, USA}
\affiliation{DAMTP, Centre for Mathematical Sciences, University of Cambridge, Wilberforce Road, Cambridge CB3 0WA, UK.}

\author{Nicol\'as Yunes}
\affiliation{Illinois Center for Advanced Studies of the Universe, Department of Physics, University of Illinois at Urbana-Champaign, Urbana, IL 61801, USA}

\begin{abstract}
We present a systematic exploration of the loss of predictivity in Einstein-scalar-Gauss-Bonnet (ESGB) gravity
in spherical symmetry.
We first formulate a gauge covariant method of characterizing the breakdown of the hyperbolicity of the equations of motion in the theory. 
With this formalism, we show that strong geodesic focusing leads to the breakdown of hyperbolicity, and the latter is unrelated to the violation of the null convergence condition.
We then numerically study the hyperbolicity of the equations during gravitational collapse for two specific ESGB gravity theories: ``shift symmetric Gauss-Bonnet gravity'' and a version of the theory that admits ``spontaneously scalarized'' black holes.
We devise a ``phase space'' model to describe the end states for a given class of initial data.
Using our phase space picture, we demonstrate that the two theories we consider remain predictive (hyperbolic) for a range of GB couplings. 
The range of couplings, however, is small, and thus, the presence of ``spontaneously scalarized'' solutions requires fine-tuning of initial data.
Our results, therefore, cast doubt as to whether scalarized black hole solutions can be realistically realized in Nature even if ESGB gravity happened to be the correct gravitational description. 
\end{abstract}
\maketitle
\section{Introduction}
The detection of gravitational waves by the LIGO/Virgo collaboration 
has allowed for new tests of general relativity (GR) 
in the dynamical and strong field regime 
\cite{TheLIGOScientific:2016src,
   Yunes:2016jcc,
   Baker:2017hug,
   Abbott:2018lct, 
   Isi:2019aib, 
   Abbott:2020jks, Okounkova:2021xjv,Perkins_2021,Nair_2019,Lyu_2022}. 
Performing \emph{model-dependent} tests of GR, however, requires accurate template waveforms computed within specific theories of gravity beyond Einstein's
\cite{Berti:2015itd,Yagi:2016jml,Yunes:2016jcc,Berti:2018cxi,Barack:2018yly}.
If the compact objects in a binary system are widely separated, then, one can use the post-Newtonian (PN) approximation to build accurate waveforms.
Waveforms built from the PN approximation already exist both in~\cite{Blanchet_Review,Blanchet:1996pi,Will:1996zj,Blanchet_2008_3PN} and outside GR~\cite{Yagi:2011xp,Sennet-2016,Shiralilou:2020gah,Shiralilou:2021mfl}. 
Near the merger, however, the PN approximation is not enough and full numerical relativity simulations are needed, again both in and outside GR. 
Although such simulations are now routinely possible within GR~\cite{Pretorius-Evolution,Baker_2006,Campanelli}, simulations outside of GR are only in their infancy~\cite{Okounkova:2017yby,Okounkova:2019dfo,Okounkova:2019zjf,Okounkova_2020,Witek:2018dmd,Silva:2020omi,Elley:2022ept,East:2020hgw,East:2021bqk,East:2022,Figueras:2021abd,CCZ4-2022,Ripley:2022cdh,Corman:2022xqg}. 

One class of theories that has received much attention is 
Einstein scalar Gauss-Bonnet (ESGB) gravity.
This theory consists of a scalar field $\phi$ 
that non-minimally couples to the Gauss-Bonnet curvature
invariant through a scalar potential $f(\phi)$ and a coupling constant $\ell$.
The scalar Gauss-Bonnet coupling appears in the low-energy limit
of heterotic string theory~\cite{Zwiebach:1985uq,Gross:1986mw,Cano:2021rey},
and, more generally,  
in effective field theories that include a real scalar field~\cite{Weinberg_2008,Kovacs_2020_PRL}.
For either case, ESGB theory parametrizes a leading
order gradient correction to the Einstein equations that involves
a scalar field~\footnote{If one were to write 
down all possible set of terms in the action that contain up to four derivatives, one can have additional terms, such 
as $\alpha(\phi) (\nabla \phi)^4$~\cite{Weinberg_2008,Kovacs_2020_PRL}, 
which can impact the dynamics of 
scalar hairy black holes (BHs) \cite{CCZ4-2022}. For simplicity,
here we set $\alpha=0$.}. 
Solutions to ESGB gravity have received much recent attention
because, for some couplings $f(\phi)$, the theory admits scalar hairy BH
(for example
\cite{1996-Kanti,Yunes_Stein:2011we,Sotiriou:2013qea,Sotiriou_2014,Ayzenberg_2014,Pani-slowly-rotating,Maselli-EdGB,Silva_2018,Doneva_2018,Doneva:2017bvd}; see
\cite{Herdeiro:2015waa} for a general 
review).
Binaries composed of scalar hairy BHs radiate scalar radiation, which impacts
the rate of inspiral, and more generally the morphology of the
radiated gravitational waves \cite{Yagi:2011xp,Shiralilou:2020gah,Lyu_2022}.
Because of this, ESGB gravity is an interesting theory to study in the
context of binary BH mergers and tests of GR with gravitational waves 
\cite{Berti:2015itd,Yunes:2016jcc,Yagi:2016jml,Barack:2018yly,Berti:2018cxi}.

All numerical relativity studies of BH mergers in ESGB gravity have relied on one of the following approaches, i) a perturbative
approach to solve the field equations, ii) a recently-constructed, strongly-hyperbolic formulations of the field equations~\cite{Kovacs_2020_PRL,Kovacs:2020ywu}
or iii) have used a ``fixing'' approach to study the dynamics~\cite{Franchini:2022ukz}.
In the perturbative approach, one solves the ESGB field equations order-by-order in $\ell^2$.
This method robustly capture the 
scalar field dynamics about BHs ~\cite{Elley:2022ept,HegadeKR:2022xij,Witek:2018dmd,Benkel:2016kcq,Benkel:2016kcq,Silva:2020omi}, 
with fairly general results obtained recently for the 
growth of monopole and dipole scalar hair~\cite{HegadeKR:2022xij,AHKR-2022-Axi-Symmetry}.
Higher-order perturbative solutions can in principle
captures nonlinear gravitational effects, 
such as the dephasing of BH binaries
due to the emission of scalar radiation \cite{Okounkova_2019,Okounkova_2020}.
But the perturbative approach is self-consistent only 
when the relative corrections to the Einstein equations 
remain ``small,'' and this is not the case at higher order
due to secular growth of uncontrolled remainders \cite{bender1999advanced,Flanagan:1996gw,Okounkova_2019}. 
The later may be cured, at least in principle, through 
techniques from multiple scale analysis and dynamical renormalization~\cite{GalvezGhersi:2021sxs}, 
but this has yet be to be applied to gravitational waveform modeling.

\begin{figure*}[thp!]
    \centering
    \includegraphics[width = 1\textwidth]{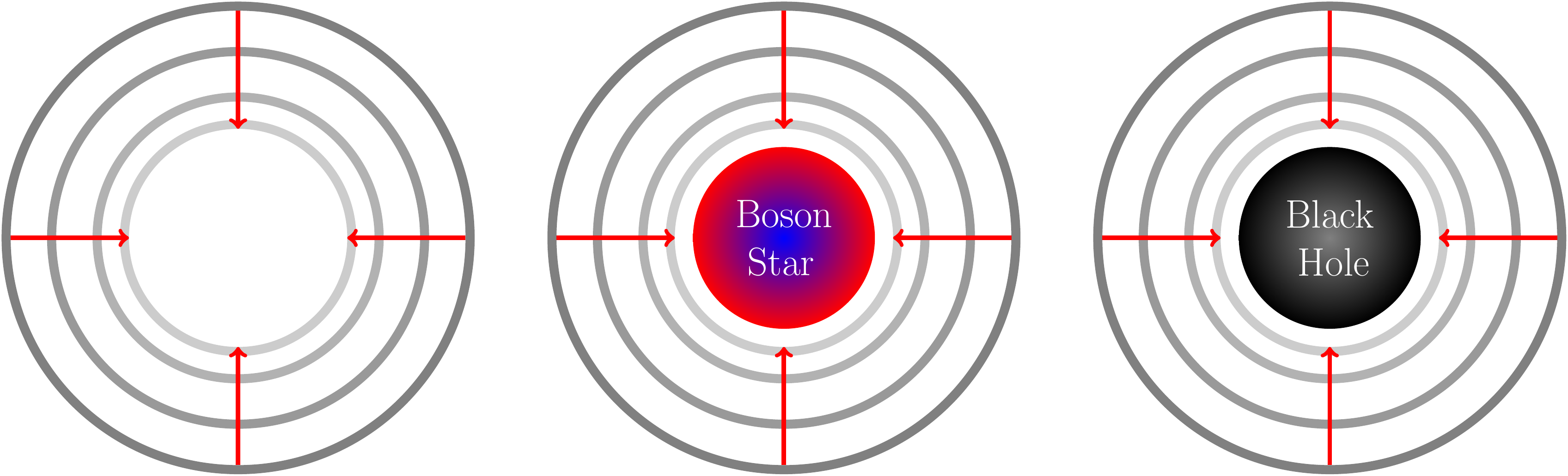}
    \caption{Cartoon depicting gravitational collapse of a shell of GB scalar field $\phi$.
    The shell starts at the outer most layer and collapses inwards as time advances, as indicated by the red arrows.
    The three different panels show the different types of initial data we study (see Sec.~\ref{sec:ID} and Table~\ref{tab:IC}).
    On the left most panel, we show the gravitational collapse of the scalar field in an otherwise flat spacetime (Gauss-Bonnet Collapse Initial Condition--GBCIC). 
    The middle panel shows the in-fall of the scalar field onto a boson star (Stable Boson Star Initial Condition--SBSCIC). The right panel
    depicts the in-fall of the scalar field onto a Schwarzschild BH
    (BH Initial Condition--BHIC).}
    \label{fig:cartoon-collapse}
\end{figure*}

In the strongly-hyperbolic 
approach, the equations of motion are solved without approximation 
(beyond that induced by the
error of the numerical discretization of the partial differential equations,
and the truncation/compactification of the computed spacetime).
This approach, however, is only ``feasible'' (i.e.~admits a well-posed set of evolution hyperbolic equations) for \emph{weakly-coupled} solutions, where
the curvature scales in the theory are large compared to the
length scale set by the Gauss-Bonnet curvature
\cite{Kovacs_2020_PRL,Kovacs:2020ywu,East:2020hgw,East:2021bqk,CCZ4-2022,Ripley:2022cdh}.
This approach avoids the secular growth of uncontrolled remainders, 
and can straightforwardly capture important nonlinear effects, such
as the dephasing of BH binaries due to the emission
of scalar radiation\cite{Flanagan:1996gw,East:2020hgw,East:2021bqk,CCZ4-2022}.
The approach, however, breaks down generically for \emph{strongly-coupled} solutions,
which can be interpreted as signalling the importance
of (unaccounted-for) higher-order gradients in the action that would arise, for
example, from the low-energy limit of a string theory 
\cite{Gross:1986mw,Cano:2021rey}. 

In this work, we present a simple, covariant explanation for why and 
ESGB breaks down in the strongly-hyperbolic approach for strongly-coupled
solutions in spherically symmetric spacetimes.
Our approach also provides diagnostics that can be used to understand if 
the corrections to Einstein's equations remain ``small'' in the perturbative approach.
We build on previous numerical work in ESGB gravity and spherical symmetry, 
which showed that the field equations can change character from hyperbolic 
to elliptic during evolution~\cite{Ripley:2019-ESGB,Ripley:2019aqj,Ripley_2020_shift_symm_PG,Ripley:2020vpk,East:2021bqk,Corelli:2022pio,Corelli:2022phw,Ripley:2022cdh}.
If the equations change character outside an event horizon, 
then the region where this breakdown occurs will be called 
a \textit{naked elliptic region} (NER),
in analogy to the concept of a naked singularity in pure general relativity.
We extend the previous studies by deriving a \emph{gauge-covariant} expression for the principal symbol of 
ESGB gravity for spherically-symmetric spacetimes 
and we show that the emergence of NERs is gauge 
covariant\footnote{Some properties of the principal symbol and characteristic 
polynomial for ESGB gravity--and other theories that have second order
equations of motion--are derived in \cite{Reall:2021voz}.}.

We then go beyond previous work by studying the mathematical and physical reasons for the emergence of NERs.
The presence of scalar hair in ESGB gravity leads to the violation of null convergence condition (NCC)~\cite{Ripley_2020_shift_symm_PG,Ripley:2019-ESGB,Ripley:2019aqj,Ripley:2020vpk}, and this has been thought to be correlated with the appearance of NERs. 
We show that NERs actually appear in regions where the NCC 
condition is \textit{not} violated.
Instead, we find that NERs appear when there is a strong focusing of null 
geodesics.
More precisely, our results indicate that the breakdown of hyperbolicity is a non-perturbative effect entering at $\mathcal{O}(\ell^{-8})$ due to strong focusing.
We provide geometric quantities that can be used to diagnose the appearance of 
NERs in the decoupling limit and in the case of full non-linear evolution.

We make the above generic statements concrete by studying the dynamics of 
two specific types of ESGB theories, classified by the choice of coupling function $f(\phi)$: 
``shift-symmetric ESGB gravity''
(sGB gravity) $f(\phi) = \phi$~\cite{Yunes_Stein:2011we} 
and a ``Gaussian'' coupling function
$f(\phi) = \left(1-\exp\left(-3\phi^2\right)\right)/6$~\cite{Doneva:2017bvd}.
In the terminology of~\cite{Elley:2022ept}, 
the shift symmetric theory represents a Type-I theory ($f'(\phi=0)\neq0$)
and the Gaussian theory represents a Type-II theory ($f'(\phi=0)=0$).
Compact objects in Type-I theories are always scalarized, while compact objects 
in Type-II theories can admit both GR solutions and scalarized solutions~\cite{Doneva:2017bvd,Silva_2018}, depending on
the compactness of the object and the type of initial data considered.

To understand the breakdown of ESGB gravity for these two choices of coupling functions,
we consider three dynamical situations in spherical symmetry, each of which provides a toy model to understand
the complicated dynamics of full $3+1$ evolution:
\begin{enumerate}
    \item Gravitational collapse of the Gauss-Bonnet 
    scalar field in an otherwise Minkowski spacetime. 
    \item In-falling Gauss-Bonnet scalar field into a 
    stable boson star in its ground state.
    \item In-falling Gauss-Bonnet scalar field into a 
    Schwarzschild BH.
\end{enumerate}
Fig.~\ref{fig:cartoon-collapse} is a cartoon that depicts these different
scenarios, which is discussed in more detail in Table~\ref{tab:IC} and Sec.~\ref{sec:ID}.

We organize the possible late-time end states for these three kinds of initial data using a ``phase-space'' diagram. 
For both the shift-symmetric and Gaussian theory,
all three classes of initial data give qualitatively similar end-states.
For sufficiently weak initial data with initial gradients much less than
$1/\ell$, the theory does not break down. 
If the initial data is sufficiently strong (in the sense that the Arnowitt-Deser-Misner (ADM) mass
of the initial data is large), and does not contain large gradients
(the smallest curvature scale is still large compared to $1/\ell$)
then the evolution is also stable and ends in the formation of a stable and large BH
In between these two end states,
there exists a ``gap'' in the phase space inside which the evolution breaks down due to the
formation of a NER. As we discuss in Sec.~\ref{sec:Numerical-Results}, 
the presence of this gap essentially
precludes the study of critical collapse in ESGB gravity. 
Such a result is consistent with the expectation that 
the theory breaks down when curvatures are large, as would
be the case when small BHs form near the threshold 
between collapse and dispersal. 

We also map out the size of the smallest possible BHs 
in the theory that form without the emergence of NERs, 
for our choices of initial data.
For the Gaussian theory, we find that the smallest possible BHs 
always lie above the the allowed range of masses that can exhibit 
spontaneous scalarization.
This implies that the phase space available for some phenomena, such as spontaneous scalarization and de-scalarization~\cite{Elley:2022ept,Silva:2020omi}, may be very narrow and might require fine tuning of initial data.
This result was hinted at in previous work~\cite{East:2021bqk} and we provide conclusive evidence by using our phase-space picture.

The rest of the paper explains all of the above results in detail and is organized as follows.
Section~\ref{sec:Field-eqs-diagnostics} describes the field equations and provide the gauge
invariant approach to study the character of ESGB gravity in spherical symmetry. 
The details of our numerical setup and our numerical results for the phase space of ESGB gravity are presented in
Sec.~\ref{sec:Numerical-Results}.
Our conclusions and directions for future work are presented in Sec.~\ref{sec:Conclusions}.
Henceforth, we use the following conventions: 
the signature of our metric is $(-,+,+,+)$,
and we use geometric units $G=1=c$.
We also introduce a fiducial length scale
$M_{\star}$ which will be used to scale physical quantities with dimensions of length. 
So, unless otherwise stated, physical quantities with the dimension of $(\text{length})^{p}$ will be assumed to be scaled with $M_{\star}^{-p}$.
\section{Field Equations and Characteristics}\label{sec:Field-eqs-diagnostics}
In this section, we begin by describing the field equations and the equations of motion for ESGB in Sec.~\ref{sec:Field-equations}.
We describe our notation in Sec.~\ref{sec:notation} and present our gauge invariant approach for calculating the principal symbol in spherically symmetric spacetimes in Sec.~\ref{sec:construction-of-principal-symobl}.
We finally analyze the principal symbol in a local null-frame and present different diagnostic tools which can be used to understand the breakdown of hyperbolicity in Sec.~\ref{sec:null-frame-analysis}.

\subsection{Field Equations}\label{sec:Field-equations}
The action for ESGB gravity is given by
\begin{equation}\label{eq:action-grav}
    S = \frac{1}{8\pi} \int d^4x \sqrt{-g} \left( 
        \frac{R}{2} 
        - 
        \frac{\left(\nabla \phi\right)^2}{2} 
        + 
        \ell^2 f(\phi)\mathcal{G}
    \right) 
    + 
    S_{\text{matter}}
    \,,
\end{equation}
The field equations derived from the above action are
\begin{align}
    \label{eq:grav-equations}
    E_{\mu\nu} 
    &:= 
    G_{\mu\nu} 
    - 
    \left(
        \nabla_{\mu} \phi \nabla_{\nu}\phi 
        -
        \frac{g_{\mu\nu}}{2} (\nabla \phi)^2
    \right) 
    \nonumber \\
    &
    + 
    2 \ell^2
    \delta^{\alpha \beta \gamma \delta }_{\kappa \sigma \rho(\mu}g_{\nu)\delta }
    R^{\kappa \sigma}{}_{\alpha \beta }\nabla^{\rho} \nabla_{\gamma} f(\phi) 
    - 
    8\pi T^{\text{matter}}_{\mu\nu} 
    =
    0
    \,,\\
    \label{eq:scalar-field-equation}
    E_{\phi} &:= \Box\phi + \ell^2\,f'(\phi)\mathcal{G} =0\,
\end{align}
where $\delta^{\alpha\beta\gamma\delta}_{\kappa\sigma\rho\mu}$ is the generalized
Kronecker delta tensor, and $T_{\mu\nu}^{\rm matter}$ is the stress-energy
tensor for other matter fields.
For the analysis presented in this section we make no assumptions about the matter
stress energy tensor beyond that it consists of only first order derivatives 
acting on the matter fields, and it describes matter fields that are minimally coupled
to the spacetime metric.
We also find it convenient to define a ``total'' stress energy tensor, which 
includes the ``massless'' piece of the Gauss-Bonnet scalar field and the 
matter stress energy tensor
\begin{equation}\label{eq:full-stress-tensor}
    T_{\mu\nu} := \left(\nabla_{\mu} \phi \nabla_{\nu}\phi -\frac{g_{\mu\nu}}{2} (\nabla \phi)^2\right) + 8 \pi T_{\mu\nu}^{\text{matter}}\,.
\end{equation}
In Sec.~\ref{sec:Numerical-Results} we use a complex scalar field $\rho$ with a mass $m_b = M_{b}/\hbar$ which admits boson-star solutions~\cite{Hawley_2000,Liebling_2017} for the matter model
\begin{subequations}
\label{eq:boson_star_action}
\begin{align}
    S_{\text{matter}} 
    &= 
    \frac{1}{16\pi} \int d^4x \sqrt{-g} \left(
        -
        \nabla^a\rho \nabla_a\rho^{*} 
        - 
        m_b^2 \rho \rho^{*}  
    \right)
    \,, \\
    E_{\rho} 
    &:=
    \Box\rho 
    - 
    m_b^2\,\rho 
    =0
    \,.
\end{align}
\end{subequations}
We briefly review some basic properties of boson stars
in Appendix~\ref{appendix:Boson-star}; 
see Refs.~\cite{Liebling_2017,Hawley_2000} for comprehensive reviews. 
We use the boson star solutions we construct
as a stand-in toy model for more realistic stars (such as neutron stars and white dwarfs).
\subsection{Gauge invariant notion of hyperbolicity for spherically symmetric spacetimes}\label{sec:gauge-inv-principal-symbol}

We now outline our derivation of a covariant
expression for the principal symbol of ESGB gravity
in spherically symmetric spacetimes.
We adopt the notation of Refs.~\cite{Papallo:2017qvl,Kovacs:2020ywu,Ripley:2022cdh} 
and we refer the reader to 
Refs.~\cite{CB-2010,Sarbach_2012,HILDITCH_2013,Kovacs:2020ywu,Ripley:2022cdh} 
for a more detailed account of the principal symbol and its relation to the 
well-posedness of the initial value problem.

We first review how the principal symbol is defined.
We consider a system of partial differential equations (PDEs) 
$E_{I}(x,u,\partial u, \partial^2 u)$, where the spacetime coordinates
are given by $x^{\mu}$, and $u^{J}$ denote the evolved fields.
The index $I \in \left(1,2,\ldots, N\right)$ is used to count the number of equations, where $N$ is the total number of fields.
Given a covector $\xi_{\mu}$, the principal symbol is defined to be \cite{CB-2010,Ripley:2022cdh,Kovacs:2020ywu,Sarbach_2012}
\begin{align}\label{eq:principal-symobl-defintion}
    P_{IJ}(\xi) := P_{IJ}^{\mu\nu}\xi_{\mu}\xi_{\nu} = \frac{\partial E_{I}}{\partial \left(\partial_{\mu}\partial_{\nu} u^{J} \right)}\xi_{\mu}\xi_{\nu}\,,
\end{align}
and as a shorthand, we will write $P\left(\xi\right)=P_{IJ}\left(\xi\right)$.
We say a covector $\xi_{\mu}$ is characteristic if it satisfies the characteristic equation
\begin{align}
    \text{det} \left( P(\xi) \right) &=0\,.
\end{align}
The system of partial differential equations $E_I$ at a spacetime point $x^{\mu}$ are said to be
\begin{enumerate}
    \item \textit{Hyperbolic}, the all solutions of the characteristic equation are real.
    \item \textit{Elliptic}, if all the solutions to the characteristic equation are imaginary.
\end{enumerate}
If there is only one dynamical field $N = 1$ then, the character of the equation can be analyzed by looking at the signature of the principal symbol. So, for a scalar equation the above definitions can be stated in the following equivalent form~\cite{CB-2010}.
Given a scalar PDE
\begin{align}
    E(x,u,\partial u, \partial^2 u) &= 0\,
\end{align} 
the PDE at a spacetime point $x^{\mu}$ is said to be
\begin{enumerate}
    \item \textit{Hyperbolic}, if the signature of the matrix $P^{\mu\nu}$ is Lorentzian, i.e. $P^{\mu\nu}$ has one negative eigenvalue and the other eigenvalues are positive. 
    \item \textit{Elliptic},  if the matrix $P^{\mu\nu}$ is positive or negative definite i.e., the eigenvalues are all positive or all negative.
\end{enumerate}
We now focus on the ESGB field equations in spherical symmetry and state 
the simplifications that can be used to calculate the principal symbol.
The propagation of the scalar field $\phi$ is governed 
by~Eq.~\eqref{eq:scalar-field-equation}.
As we see from Eq.~\eqref{eq:scalar-field-equation},
the Gauss-Bonnet scalar $\mathcal{G}$ contains second derivatives of the metric.
We show that for a spherically symmetric spacetime one can use 
the tensor equations of motion to trade the second derivatives of the metric in 
$\mathcal{G}$ for second derivatives of the scalar field. 
This means that we can rewrite the scalar field equation to take the form
\begin{align}
    E_{\phi} 
    = 
    \Box \phi 
    + 
    \ell^2 f' \, \mathcal{G}[\partial^2 \phi,\partial \phi, \phi]
    \,.
\end{align}
After this simplification is achieved, we can
focus on the above equation and calculate the principal symbol using Eq.~\eqref{eq:principal-symobl-defintion} for the gauge-invariant scalar
field $\phi$. 

We stress that this approach relies heavily on the spherical symmetry 
of the spacetime. Outside of spherical symmetry, one must generally contend
with gauge degrees of freedom, which complicate the analysis of the characteristics;
for more discussion see \cite{Papallo:2017qvl,Kovacs_2020_PRL,Kovacs:2020ywu,Reall:2021voz}.

\subsubsection{Notation}\label{sec:notation}
Here we set the notation we use to derive the 
principal symbol and understand its properties in spherical symmetrically symmetric
spacetimes.
We decompose the 4-D metric $g_{\mu\nu}$ as~\cite{Maeda_2008}
\begin{equation}\label{eq:spherical-metric}
    ds^2 = \alpha_{ab} du^a\, du^b + r^2(u^a) \Omega_{AB} d\theta^A d\theta^B
\end{equation}
The function $r(u^0,u^1)$ measures the proper radius of the 2-sphere and 
$\Omega_{AB}$ is the standard metric on the 2-sphere.
The above metric is the most general metric for a spherically symmetric spacetime. 
To simplify our analysis we will use the following notation:
upper case Latin letters $\left(A,B,\ldots \right)$ 
will be used to represent indices on the 2-sphere, 
lower case Latin letters $\left(a,b,\ldots \right)$ 
will be used to represent indices
``perpendicular'' to the 2-sphere in the ``t-r'' plane,
and lower case Greek letters $\left(\mu,\nu,\ldots \right)$
will be used for general four dimensional indices.
We will use $\nabla_{\mu}$ to denote the 4-D covariant derivative and 
$D_{a}$ to denote the 2-D covariant derivative compatible with $\alpha_{ab}$.
The Christoffel symbol and Riemann tensor for the metric of
Eq.~\eqref{eq:spherical-metric} are listed in Appendix~\ref{appendix:curvature-components}.

To simplify our calculations, we find it useful to introduce the following linear operators (here $F$ is any scalar function)
\begin{align}
    \label{eq:def-lambda-stf}
    \lambda^{<ab>}[F] &:= D^a D^b F - \frac{1}{2} \alpha^{ab} D^2F  \,,  \\
     \label{eq:def-lambda}
    \lambda[F] &:= D^2F.
\end{align}
The first operator is a symmetric trace free (STF) operator and $D^2 := D^a D_a$ is the d'Alembertian.
We will also use the following notation to denote the two-dimensional 
STF operation and trace on a general tensor $\Sigma_{ab}$
\begin{align}
    \Sigma_{<ab>} &:= \Sigma_{(ab)} - \frac{1}{2} \alpha_{ab} \Sigma_2\,,\\
    \label{eq:2-trace-def}
    \Sigma_2 &:= \alpha^{ab} \Sigma_{ab}\,.
\end{align}
We denote the four dimensionsal trace with
\begin{align}
    \Sigma_{4} := g^{\mu\nu}\Sigma_{\mu\nu} &= \alpha^{ab}\Sigma_{ab} + \frac{\Omega^{AB}}{r^2}\Sigma_{AB}\,, \\
    &= \Sigma_2  + \frac{\Omega^{AB}}{r^2}\Sigma_{AB}
    \,
    ,
\end{align}
and the 
difference between a 2-D trace and a 4-D trace by $\Tilde{\Sigma}$ 
\begin{align}
    \label{eq:tilde-def}
    \Tilde{\Sigma} &:= \frac{\Omega^{AB}}{r^2}\Sigma_{AB} = \Sigma_4 -\Sigma_2 \,.
\end{align}
Finally, we introduce three scalar functions
\begin{align}
    \label{eq:sigma-def}
    \sigma &:= (Dr)^2-1 = - \frac{2 M}{r} \,,\\
    \label{eq:mu-def}
    \mu &:= r - 8\ell^2(D^cr)(D_cf)\,,\\
    \label{eq:pi1-def}
    \pi_1 &:= \frac{96 \sigma (f')^2}{r^2\mu} = -\frac{192 M (f')^2}{r^3 \mu} \,,
\end{align}
where the scalar function $M(u^0,u^1)$ is the quasi-local Misner-Sharp mass~\cite{Maeda_2008}.
\subsubsection{Projection of the equations of motion}

We now outline our derivation of the principal symbol, 
leaving details to Appendix~\ref{appedix:proj-eom}.
We first begin by studying the projections $E_{ab}$ and $E_{AB}$ of the gravitational field equations $E_{\mu\nu}$ of Eq.~\eqref{eq:grav-equations}. 
The tensor $E_{ab}$ allows us to solve for $\lambda_{<ab>}[r]$ and $\lambda[r]$ as functions of $\lambda_{<ab>}[f]$ and $\lambda[f]$ respectively,
where $f:=f\left(\phi\right)$ is shorthand
for the scalar Gauss-Bonnet coupling function.
The final solutions are given in Eqs.~\eqref{eq:lambda-stf-sol} and \eqref{eq:lambda-sol}.

Next, we study the projection of the gravitational tensor equations
of motion on its spherical indices ($E_{AB}$),
which allows us to obtain the two-dimensional Ricci scalar $\mathcal{R}(\alpha_{ab})$ as a function of $\lambda_{<ab>}[f]$ and $\lambda[f]$.
The solution for $\mathcal{R}$ is given in Eq.~\eqref{eq:ricci-2-sol}.

For the benefit of the reader, we present here some of the final
expressions that will be important below, namely
\begin{align}
    \label{eq:lambda-stf-reduced}
    \lambda_{<ab>}[r] &= \frac{4\ell^2\sigma}{\mu}\lambda_{<ab>}[f] - \frac{r^2}{2\mu} T_{<ab>} \,,\\
    \label{eq:lambda-r-reduced}
    \lambda[r] &= \frac{\left(4\ell^2 \lambda[f] -1\right) \sigma}{\mu} + \frac{r^2 T_2}{\mu}\,,\\
    \mathcal{R} &= \frac{16\ell^2}{\mu} \lambda_{<cd>}[f]\lambda^{<cd>}[r] - \frac{8\ell^2}{\mu}\lambda[f]\lambda[r] + \frac{2 \lambda[r]}{\mu} \nonumber\\
    &-\frac{r\Tilde{T}}{\mu}\,,\\
    \label{eq:GB-v1}
    \mathcal{G} &= \frac{12}{r^2} \lambda[r]^2 - \frac{24}{r^2}\lambda_{<ab>}{[r]}\lambda^{<ab>}[r]  \nonumber \\
    &- \frac{8}{\mu}\lambda^{<ab>}[r] T_{<ab>} - \frac{8 T_2}{\mu}\lambda[r] + \frac{4 \sigma}{\mu r^3} \Tilde{T}\,,
\end{align}
where, $T_{<ab>}$ denotes the STF part of the stress energy tensor the functions  $T_2$, $\Tilde{T}$, $\sigma$ and $\mu$ are
defined in Eqs.~\eqref{eq:2-trace-def}, \eqref{eq:tilde-def},
\eqref{eq:sigma-def} and \eqref{eq:mu-def} respectively.
The first three equalities come from the field equations for the metric tensor, 
while the last one can be computed from the definition of the Gauss-Bonnet invariant
and the above expressions. 

One can also think of the above equations 
as providing a relation between derivatives of the metric functions and derivatives of the scalar field, 
which one can now use in the equation of motion for the scalar field. 
More precisely, the scalar field equation of motion can be written as
\begin{align}
    E_{\phi} &= \Box{\phi} + \ell^2 f' \mathcal{G} = 0\,, \nonumber\\
    \label{eq:E-phi-v1}
    &= D^2 \phi + \frac{2}{r} D^ar D_a \phi + \ell^2 f' \mathcal{G}[f] = 0
    ,
\end{align}
where in the second line $\mathcal{G}[f]$ is a function of derivatives of $f(\phi)$, when one 
substitutes Eqs.~\eqref{eq:lambda-stf-reduced} and~\eqref{eq:lambda-r-reduced} into Eq.~\eqref{eq:GB-v1}. 
The final expression for $\mathcal{G}[f]$ is long and un-illuminating, so we present it in Eq.~\eqref{eq:gb-func-f}.

\subsubsection{Expression for the principal symbol}\label{sec:construction-of-principal-symobl}

Given Eq.~\eqref{eq:E-phi-v1}, we 
we can now obtain the principal symbol for the
scalar degree of freedom 
(more details are given in Appendix.~\ref{appendix:pab-construction}). 
We start by looking at the scalar field equations of motion [Eq.~\eqref{eq:E-phi-v1}],
where $\mathcal{G}$ in Eq.~\eqref{eq:GB-v1} is now viewed as a function which depends on $f$ through $\lambda_{<ab>}[r]$ in Eq.~\eqref{eq:lambda-stf-reduced} and $\lambda[r]$ in Eq.~\eqref{eq:lambda-r-reduced}.
The principal symbol is therefore given by
\begin{align}
    \mathcal{P}\left[E_{\phi}\right] &= \mathcal{P}\left[D^aD_a \phi\right] + \ell^2 f' \mathcal{P}\left[\mathcal{G}[f]\right]
\end{align}
where we have ignored the lower order terms. 
The first term in the above equation is given by
\begin{align}
    P\left[D^aD_a \phi\right] = \alpha^{ab}\xi_{a}\xi_{b}\,,
\end{align}
while the second term $P\left[\mathcal{G}[f]\right]$ is calculated in Eq.~\eqref{eq:pg-simplified}.
Using Eq.~\eqref{eq:pg-simplified}, 
we see that the principal symbol can be written as in Eq.~\eqref{eq:pab-ricci-appendix}
\begin{align}
    \label{eq:pab-ricci}
    P_{ab} &= \alpha_{ab}\left[1 + \pi_1 \ell^4 \left(\lambda[r] - \frac{r^2 }{3\mu}T_2 \right)\right] \nonumber \\
    &+ \pi_{1} \ell^4 r \left( R_{<ab>} - \frac{r}{3\mu} T_{<ab>} \right)\,,
\end{align}
where $R_{<ab>}$ is the two dimensional STF form of the Ricci tensor (see Eq.~\eqref{eq:Ricci-simplified}) 
and $T_{<ab>}$ is the two dimensional STF form of the full stress energy tensor, defined in Eq.~\eqref{eq:full-stress-tensor}.
For ease of numerical implementation, we also provide two other equivalent forms of the principal symbol in Eqs.~\eqref{eq:pab-ricci-lf-appendix} and \eqref{eq:pab-ricci-lf-labf-appendix} in terms of the derivatives of the scalar field.

The above equation is gauge invariant in the following sense. This equation describes the principal symbol for the scalar degree of
freedom in ESGB gravity in spherical symmetry, which is the
only gauge-invariant dynamical degree of freedom for the
theory in such spacetimes (outside of the matter fields, which again we assume
are minimally coupled to the spacetime metric and
do not coupled with derivatives to $\phi$).
Our derivation hinged on the fact that we could replace second
derivatives of the metric functions $\alpha_{ab}$ and $r$
with second derivatives of $\phi$.
This is possible because there are effectively no tensor degrees of freedom in spherical symmetry.
\subsection{Analysis in a local null frame}\label{sec:null-frame-analysis}

Solutions to ESGB gravity are well-known to typically contain regions inside which the NCC is violated
~\cite{Ripley:2019-ESGB,Ripley:2019aqj,Ripley_2020_shift_symm_PG,Ripley:2020vpk}.
The NCC is a crucial ingredient in many classical results on the properties of BHs, most notably the area theorem
~\cite{Hawking:1971tu,Hawking:1973uf}.
During dynamical evolution, the area of the BH is known to decrease in ESGB theory; heuristically, the growth of the scalar
hair around the BH extracts energy from it, which
forces the area to shrink in size\footnote{We note that although the area of the BH decreases if the strength of initial data is large, this does not mean that the second law of BH fails to hold. 
For weak perturbations of a stationary BH there exists a prescription to calculate the BH entropy in ESGB gravity where the second law holds ~\cite{Hollands:2022fkn}.}~\cite{Ripley:2019-ESGB}.
However, there need not be any connection between violations of the NCC
and the hyperbolicity of a general theory of gravity. 
For example, a ghost field has hyperbolic equations of motion,
and solutions to the theory violate the NCC as long as
$\nabla_{\mu}\phi\neq0$.
Nevertheless, one may suspect that, for theories
like ESGB gravity, which are supposed to capture leading-order 
effective corrections to the Einstein equations,
there could be some connection between the breakdown of the 
theory and large violations of the NCC.
Below we show that this suspicion is unfounded: there is no
actually connection between violations of the NCC and the
breakdown of the full equations of motion in ESGB gravity,
at least in spherically symmetric spacetimes.

To show this, we decompose the spacetime into null components.
In spherical symmetry, there are two preferred null frame vectors
that are orthogonal to the two spheres that foliate the spatial slices 
(surfaces with constant areal radius $r$).
Consider a local null frame with outgoing null vector $k^{a}$ and ingoing null vector $l^a$. 
The metric in this local frame is given by
\begin{equation}
    \alpha_{ab} = -2 l_{(a}k_{b)} \,,\, l^a k_{a} = -1\,.
\end{equation}
The components of the matrix $P_{ab}$ of Eq.~\eqref{eq:pab-ricci} 
in this local frame are then given by
\begin{equation}
    \mathbb{P} := 
    \begin{pmatrix}
    P_{ab}k^a k^b & P_{ab}l^a k^b \\
    P_{ab}l^a k^b & P_{ab}l^a l^b
    \end{pmatrix}\,,
\end{equation}
where
\begin{align}
    P_{ab}k^a k^b &= \pi_1 \ell^4 r \left(R_{kk} - \frac{r}{3\mu} T_{kk} \right) := \pi_1 \ell^4 r B_{kk}\,,\\
    P_{ab}l^a k^b &= - \left[1 + \pi_1 \ell^4 \left(\lambda[r] - \frac{r^2}{3\mu} T_2\right)\right]\,, \\
    P_{ab}l^a l^b &= \pi_1 r \left(R_{ll} - \frac{r}{3\mu} T_{ll} \right) := \pi_1  \ell^4 r B_{ll}\,,
\end{align}
and the determinant is
\begin{equation}\label{eq:detP-null-frame}
    \text{det}\, \mathbb{P} = \pi_1^2\, \ell^8\, r^2 B_{ll} B_{kk} - \left[1 + \pi_1 \ell^4 \left(\lambda[r] - \frac{r^2}{3\mu} T_2\right)\right]^2\,.
\end{equation}
In the above expressions, we have used $B_{kk}$ as a shorthand for $B_{ab}k^ak^b$ and similar shorthands are used for other quantities.
As we discuss in Sec.~\ref{sec:breakdown_gradient_expansion_diagnostic},
the transition from hyperbolic to elliptic equations occurs when $\detP = 0$.
If $\detP<0$, then the equation of motion for the scalar field is hyperbolic. 
The second term in Eq.~\eqref{eq:detP-null-frame} is negative definite,
and arises from the trace of the principal symbol of Eq.~\eqref{eq:pab-ricci}.
The first term may or may not be positive definite, and it 
comes from the STF part of the principal symbol~[Eq.~\eqref{eq:pab-ricci}].
We can then think of the failure of hyperbolicity as arising from
``shear'' terms in the principal part of the equations of motion
(because shear is typically generated by STF parts of tensors).
We then conclude that the character of the equation of motion for the 
scalar field changes from hyperbolic to elliptic when
\begin{equation}\label{eq:change-condition}
    \pi_1^2\, \ell^8\, r^2 B_{ll} B_{kk} \geq \left[1 + \pi_1 \ell^4 \left(\lambda[r] - \frac{r^2}{3\mu} T_2\right)\right]^2\,,
\end{equation}
which is a sufficient condition for the loss of hyperbolicity.
From Eq.~\eqref{eq:change-condition}, we see that
$B_{kk}$ and $B_{ll}$ have to be of the same sign for the change in character to occur.

We can relate $B_{kk}$ and $B_{ll}$ to the NCC to 
understand the physical significance of Eq.~\eqref{eq:change-condition}.
Rewriting some terms in this equation with the Misner-Sharp mass
function of Eq.~\eqref{eq:pi1-def}, we ultimately obtain
\begin{align}
    \label{eq:BllBkk-v1}
    B_{ll} B_{kk} \geq \frac{r^4 \mu^2}{ M^2 \ell^8 \left[192 (f')^2 \right]^2}
    \left[1 + \pi_1 \ell^4 \left(\lambda[r] - \frac{r^2}{3\mu} T_2\right)\right]^2
    .
\end{align}
Before we proceed further, let us note that obtaining necessary conditions for the failure of hyperbolicity for the field equations[Eq.~\eqref{eq:grav-equations}-\eqref{eq:scalar-field-equation}] would need a more general analysis such as the one carried out in~\cite{Kovacs:2020ywu,Reall:2021voz}.
Therefore, we caution the reader that the inequality derived in Eq.~\eqref{eq:BllBkk-v1} is only a sufficient condition and must be used as a diagnostic for the loss of hyperbolicity in spherical symmetry. If this inequality is not satisfied then the equations are not necessarily hyperbolic. Nevertheless, we find that this condition is a good diagnostic in numerical simulations (see Sec.~\ref{sec:Numerical-Results}). 

Let us now provide a better intuitive understanding of the inequality derived above in Eq.~\eqref{eq:BllBkk-v1}.
To do this, we first expand the above inequality 
\begin{align}
    \label{eq:BllBkk-v2}
    B_{ll} B_{kk} &\geq \frac{1 }{r^2} \left( \lambda[r] - \frac{r^2}{3 \mu} T_2\right)^2 + \frac{2}{\pi_1 \ell^4 r^2} \left( \lambda[r] - \frac{r^2}{3 \mu} T_2\right) \nonumber \\
    & + \frac{1}{\pi_1^2 \ell^8 r^2}\,.
\end{align}
One is typically interested in ESGB when the coupling constant is small. When $\ell$ is small, the dominant contribution on the right hand side of the above inequality is the last term, which scales as $\ell^{-8}$.

One can simplify the left hand side by noting that, to leading order in $\ell$, the functions $B_{ll}$ and $B_{kk}$ are
\begin{align}
    B_{kk} &= R_{kk} - \frac{r}{3\mu} T_{kk} = \frac{2}{3} R_{kk} + \mathcal{O}(\ell^2)\\
    B_{ll} &= R_{ll} - \frac{r}{3\mu} T_{ll} = \frac{2}{3} R_{ll} + \mathcal{O}(\ell^2)
\end{align}
where we used the expression for $\mu$ in Eq.~\eqref{eq:mu-def} and the fact that to leading order in $\ell$ the gravitational equations of motion are those in GR, $R_{<ab>} = T_{<ab>} + \mathcal{O}(\ell^2)$. 
Inserting this expansion into Eq.~\eqref{eq:BllBkk-v2}, we see that for small $\ell$ we can rewrite the inequality as 
\begin{align}\label{eq:RkkRll-v1}
    R_{kk}R_{ll} + \mathcal{O}(\ell^2) &\geq \frac{1 }{r^2} \left( \lambda[r] - \frac{r^2}{3 \mu} T_2\right)^2 \nonumber \\
    &+ \frac{2}{\pi_1 \ell^4 r^2} \left( \lambda[r] - \frac{r^2}{3 \mu} T_2\right) + \frac{1}{\pi_1^2 \ell^8 r^2}\,,\\
    \label{eq:RkkRll-v2}
    &\geq \frac{r^6}{ M^2 \ell^8 \left[128 (f')^2 \right]^2}  +  \mathcal{O}(\ell^{-6})\,,
\end{align}
where in the last line, we expanded $\pi_1$ using Eq.~\eqref{eq:pi1-def}.
From the above equation, we see that for small $\ell$, the equations are non-hyperbolic in regions where the product $R_{kk}R_{ll}$ is positive and exceeds the inequality derived above.
Therefore, just the violation of the outgoing or ingoing NCC (i.e.~the violation of the
$R_{kk}\geq0$ and $R_{ll}\geq0$ inequalities) does not necessarily lead 
to the breakdown of the equations of motion. 
Instead, we see that the equations become non-hyperbolic when there is 
strong geodesic focusing, i.e.~when $R_{kk}R_{ll} \sim \mathcal{O}(\ell^{-8})$.
We will see in Sec.~\ref{sec:Numerical-Results} (see also Fig.~\ref{fig:nec-non-violation})
that our numerical simulations lose their hyperbolic character precisely when such strong 
geodesic focusing occurs. 

Finally, we discuss what the above results mean in the context of the decoupling 
analysis employed for
example in Refs.~\cite{Elley:2022ept,HegadeKR:2022xij,Witek:2018dmd,Benkel:2016kcq,Benkel:2016kcq,Silva:2020omi}.
In the decoupling approach, one assumes that small scalar Gauss-Bonnet 
perturbations remain small during dynamical evolution of the initial data. 
For a sufficiently small duration of time, this assumption holds true.
As time advances, however, the system may evolve into a strongly gravitating one, 
even if the initial data was weak. When this occurs, the strong focusing of geodesics
may lead to the satisfaction of Eqs.~\eqref{eq:BllBkk-v1} and \eqref{eq:RkkRll-v2}, which, in turn,
will force the evolution equations to lose hyperbolicity and become ill-posed.
Therefore, at least in spherical symmetry, one diagnostic that could be 
tracked to see if the evolution equations fail would be that determined by 
Eqs.~\eqref{eq:BllBkk-v1} and \eqref{eq:RkkRll-v2}.
These equations can be evaluated on the background GR solution itself 
to see how strong the gravitational corrections are.
For a full non-linear evolution, one can directly track Eq.~\eqref{eq:detP-null-frame}.
As we have mentioned before, outside spherical symmetry one needs to worry about gravitational 
degrees of freedom and study the full principal symbol~\cite{Kovacs:2020ywu,Kovacs_2020_PRL},
which couples gravitational a scalar degrees of freedom~\cite{Reall:2021voz}. 
Therefore, obtaining a simple formula to diagnose the breakdown, 
such as the one given in Eq.~\eqref{eq:RkkRll-v2}, 
might be challenging outside of spherical symmetry.

\section{Numerical Experiments}\label{sec:Numerical-Results}

In this section we describe the results from numerically simulating spherically-symmetric gravitational collapse in ESGB gravity for the shift-symmetric theory and a Gaussian coupling function.
In Sec.~\ref{sec:ID}, we describe the details of our numerical setup and of the initial data we use in our numerical simulations.
We describe the different diagnostics we use to track the breakdown of hyperbolicity in our numerical simulations in Sec.~\ref{sec:breakdown_gradient_expansion_diagnostic}. 
In Sec.~\ref{sec:kanti-bound} we provide a brief description of static BH solution in ESGB theories and present the problems which occur as the size of the BHs get smaller and the curvature scales increase.

\begin{table*}[t]
    \centering
    \begin{tabular}{|c|c|c|c|}
    \hline
    Name & Complex Scalar Field Profile & Gauss-Bonnet Scalar Field Profile & Initial Excision Position \T\B \\
    \hline
    GBCIC & None & $\phi_{\text{bump}}(A,r_l,r_u)$ & $r=0$ \T\B \\
    \hline
    BHIC & None & $M_{BH}$; $\phi_{\text{bump}}(A,r_l,r_u)$ & $r = M_{BH}$ \T \B \\
    \hline
    SBSCIC &  $\rho_0(r)$; Boson star is in its ground state & $\phi_{\text{bump}}(A,r_l,r_u)$ & $r = 0$ \T \B \\
    \hline
    \end{tabular}
    \caption{Families of initial data used in our numerical simulations. The acronyms GBCIC, BHIC and SBSIC stand for Gauss-Bonnet Collapse Initial Condition, BH Initial Condition and Stable Boson Star Initial Condition respectively.
    See Sec.~\ref{sec:ID} for more information.}
    \label{tab:IC}
\end{table*}

\subsection{Numerical Setup}\label{sec:ID}
We first briefly describe our numerical setup.
Our code closely follows the setup of Ref.~\cite{Ripley_2020_shift_symm_PG}.
We work in Painlev\'{e}-Gullstrand (PG) coordinates, with the line element
\begin{equation}
    ds^2 = -\alpha(t,r)^2 dt^2 + (dr + \alpha(t,r)\zeta(t,r)dt)^2 + r^2 d\Omega^2\,,
\end{equation}
where $d\Omega^2=d\vartheta^2+\sin^2\vartheta d\varphi$ is the metric of a unit sphere.
Introducing the following auxiliary variables
\begin{align}
    Q(t,r) &:= \partial_r \phi(t,r)\,, \\
    P(t,r) &:= \frac{1}{\alpha(t,r)} \partial_t \phi(t,r) - \zeta(t,r) Q(t,r)\,,\\
    \rho(t,r) &:= \rho_1 + i\, \rho_2 \,,\\
    Q_{1,2}(t,r) &:= \partial_r \rho_{1,2} \,,\\
    P_{1,2}(t,r) &:= \frac{1}{\alpha(t,r)} \partial_t \rho_{1,2}(t,r) - \zeta(t,r) Q_{1,2}(t,r)
\end{align}
the equations of motion 
[Eqs.~\eqref{eq:grav-equations}-\eqref{eq:scalar-field-equation}]
schematically take the form
\begin{align}
    \label{eq:constraint-M}
    \partial_r \zeta - \mathcal{F}_{\zeta}(\Vec{v}) &= 0  \,,\\
    \label{eq:constraint-alpha}
    \frac{\partial_r \alpha}{\alpha} - \mathcal{F}_{\alpha}(\Vec{v}) &=0 \,,\\
    \label{eq:evoution-P}
    \partial_t P - \mathcal{F}_{P}(\alpha,\Vec{v}) &=0 \,,\\
    \label{eq:evolution-Q}
    \partial_t Q - \partial_r\left[\alpha \left(P +\zeta Q \right)\right] &=0 \,,\\
    \label{eq:evolution-phi}
    \partial_t \phi - \alpha \left(P +\zeta Q \right) &=0\,,\\
    \partial_t P_{1,2} - \mathcal{F}_{P_{1,2}}(\alpha,\zeta,\rho_{1,2}) &=0\,,\\
    \partial_t Q_{1,2} -\partial_r\left[\alpha \left(P_{1,2} +\zeta Q_{1,2} \right)\right] &=0\,,\\
    \label{eq:evolution-rho-12}
    \partial_t \rho_{1,2} - \alpha \left(P_{1,2} +\zeta Q_{1,2} \right) &=0\,,
\end{align}
where we have introduced the vector
\begin{equation}
    \Vec{v} = \left(\zeta,P,P',Q,Q',\phi,P_1,Q_1,P_2,Q_2,\rho_1,\rho_2\right) \, .
\end{equation}
The full expressions for these equations of motion
(modulo the presence of the matter field $\rho$)
can be found in Appendix C of~\cite{Ripley_2020_shift_symm_PG}.

The initial data we use are summarized in Table~\ref{tab:IC}.
The rescaled bump function is given by 
\begin{equation}\label{eq:bump}
    \phi_{\text{bump}}(r) = \begin{cases}
    0\, & r\leq r_l \\
    \phi_0(r,A,r_l,r_u) & r_l<r<r_u \\
    0\, & r\geq r_u
    \end{cases}
\end{equation}
where,
\begin{equation}
     \phi_0(r,A,r_l,r_u) = A (r-r_l)^2 (r - r_u)^2 e^{\left(-\frac{1}{r - r_l} - \frac{1}{r_u -r} \right)}\,.
\end{equation}
Given the profile for the scalar field $\phi$ we obtain
the initial value for the variable $Q$ by differentiating the above profile. 
To initialize the $P$ variable, we use an approximately ingoing profile
\begin{equation}\label{eq:ingoing-ics}
    P(0,r) = Q(0,r) + \frac{\phi(0,r)}{r}\,.
\end{equation}
To initialize the complex scalar field, we either set it to zero everywhere,
or use a stable boson star profile in its ground 
state~\cite{Colpi_Shapiro_Wasserman_1986,Hawley_2000,Liebling:2012fv}, 
depending on type of initial data described in 
Table~\ref{tab:IC}.
We review the solution spectrum and how we obtain the boson star initial data in Appendix~\ref{appendix:Boson-star}.

For the Gauss-Bonnet Collapse Initial Condition (GBCIC) case and the Stable Boson Star Initial Condition case (SBSCIC), the initial excision position is at $r=0$ and we use the following regularity conditions at the origin
\begin{align}
\label{eq:bcs-alpha-fs}
    \left. \partial_r \alpha \right|_{r=0} &= 0 \,,\\
    \left. \zeta \ \right|_{r=0} &= 0 \,,\\
    \left. \partial_r P \right|_{r=0} &= 0 \,, \\
    \left. Q \right|_{r=0} &= 0 \,,\\
    \label{eq:bcs-phi-fs}
    \left. \partial_r \phi \right|_{r=0} &= 0\,, \\
    \left. \partial_r P_{1,2} \right|_{r=0} &= 0 \,, \\
    \left. Q_{1,2} \right|_{r=0} &= 0 \,,\\
    \label{eq:bcs-rho-fs}
    \left. \partial_r \rho_{1,2} \right|_{r=0} &= 0\,.
\end{align}
For the BH Initial Data (BHIC) case, we excise the grid inside the apparent horizon (AH), which is located at $\zeta = 1$, and the initial excision position is set to be at $r_{ex} = M_{\text{BH}}$.
We also set the shift $\zeta$ and lapse $\alpha$ to their GR values at the initial excision position
\begin{align}
    \zeta(0,r_{\text{ex}}) & = \sqrt{\frac{2M}{r_{\text{ex}}}}\,,\\
    \alpha(0,r_{\text{ex}}) & = 1\,.
\end{align}
This is a valid initial condition as it ensures that the support of the Gauss-Bonnet scalar is outside the initial AH. This can be achieved by controlling $r_l$ and $r_u$.
At the outer boundary, we set outgoing wave boundary conditions.

We briefly describe the physical motivation behind the three types of initial data summarized in Table~\ref{tab:IC}, as follows:
\begin{enumerate}
    \item GBCIC collapses the scalar field $\phi$, with a fixed position determined by $r_l$ and $r_u$~\eqref{eq:bump}. The physical situation explored by studying GBCIC is similar to the study of critical collapse in GR~\cite{Choptuik:1992jv}, except that now the phase space of possible end states includes evolution to NERs.
    \item SBSCIC studies the ``dynamical stability'' of a non-BH compact object, when perturbed by a Gauss-Bonnet scalar field.
    \item BHIC analyzes the situation where the spacetime has a BH at $t=0$. We study the effect of a perturbation by the Gauss-Bonnet scalar with a fixed $r_l$ and $r_u$. We also use this initial data to find the smallest possible BH in these theories.
\end{enumerate}
In all cases, we follow the evolution to determine whether the end state is a BH, a boson star, flat space or a NER. 

We next describe the numerical schemes used to evolve our equations.
We solve the constraint equations~\eqref{eq:constraint-M}-\eqref{eq:constraint-alpha} using Heun's method, a second order integration method. 
To solve the evolution equations [Eqs.~\eqref{eq:evoution-P}-\eqref{eq:evolution-rho-12}], 
we use the method of lines with a second-order-accurate finite difference stencil to 
discretize the spatial derivatives. 
We evolve the discretized 
set of equation using an RK4 time integration method.
We provide further details about the code and present our convergence results in Appendix~\ref{appendix:Convg}.

Finally, we note that we have not derived a 
rigorous mathematical proof for the local existence of for the coupled systems Eqs.~\eqref{eq:constraint-M}-\eqref{eq:evolution-rho-12}.
A local existence result has been obtained for the ESGB field equations 
[Eqs.~\eqref{eq:grav-equations}-\eqref{eq:scalar-field-equation}]
in a modified harmonic formulation in Refs.~\cite{Kovacs:2021lgk,Kovacs:2020ywu}, 
but we make use of a different formulation.
Nevertheless, those results combined with the stability and convergence of our numerical code provides strong hints for the local existence result in PG coordinates [Eqs.~\eqref{eq:constraint-M}-\eqref{eq:evolution-rho-12}].
\subsection{Diagnostics and breakdown of gradient expansion 
\label{sec:breakdown_gradient_expansion_diagnostic}}

In this section we review how we diagnose the breakdown 
of the hyperbolicity in our numerical code.
We calculate the principal symbol using Eq.~\eqref{eq:pab-v1}. We then calculate the characteristic speeds $c_{\pm}$.
Given a characteristic covector $\xi_{a} = \left(\xi_t,\xi_r\right)$, 
the characteristic speed in spherical symmetry is defined by
\begin{equation}
    c := - \frac{\xi_t}{\xi_r}\,.
\end{equation}
where the $\xi_a$ satisfy the characteristic equation~\cite{Ripley:2022cdh}
\begin{align}
    \det\left[P^{ab}\xi_{a}\xi_{b}\right]
    &= 0  \\
    \implies P^{tt}\xi_t^2 + 2 P^{tr} \xi_t\xi_r + P^{rr} \xi_r^2 &=0\\
    \implies P^{tt} c^2 - 2 P^{tr} c + P^{rr} &=0\,.
\end{align}
This gives us
\begin{equation}\label{eq:cpm}
    c_{\pm} = \frac{1}{P^{tt}}\left( P^{tr} \pm \sqrt{\mathcal{-D}}\right)\,
\end{equation}
where
\begin{equation}\label{eq:determinant}
    \mathcal{D} := P^{tt} P^{rr} - (P^{tr})^2
    ,
\end{equation}
is the determinant of the (contravariant) principal symbol.
The characteristic speeds ($c_{\pm}$) for GR in PG coordinates are given by $c_{\pm} = -\alpha\left(\zeta \pm 1 \right)$. Therefore, in GR, the system of equations [Eqs.~\eqref{eq:evoution-P}-\eqref{eq:evolution-phi}] is always hyperbolic. 
This, however is not the case in ESGB gravity~\cite{Ripley_2020_shift_symm_PG,Ripley:2019-ESGB,Ripley:2019aqj,Ripley:2020vpk}.
During gravitational collapse, we track the determinant in Eq.~\eqref{eq:determinant} and the expansion of null congruences $\Theta = 1-\zeta$. 
The AH is located at $\Theta = 0$,
(in practice we find that the AH lies
close the ``sound horizon'', where the outoing scalar characteristic speed is zero) \cite{Ripley:2019aqj}.
If the determinant $\mathcal{D} $ is greater than zero before the formation of an AH, we quit the simulation, since this signals a breakdown of hyperbolicity, and any subsequent evolution would crash the simulation because of exponentially growing modes.
We also track the determinant after the formation of an AH and excise the region where $\mathcal{D}>0$. If this excision region moves outside the AH, we quit the simulation.

As a consistency check we also track the determinant of the principal symbol in the null frame [Eq.~\eqref{eq:detP-null-frame}].
We use the following null-frame
\begin{align}
    k_{\mu} &= \left(\frac{\alpha(1+\zeta)}{\sqrt{2}} , \frac{1}{\sqrt{2}},0,0\right)\\
    l_{\mu} &= \left(\frac{\alpha(1-\zeta)}{\sqrt{2}} , -\frac{1}{\sqrt{2}},0,0\right)\,
\end{align}
to calculate the determinant.

\begin{figure*}[t!]
    \centering
     \includegraphics[width = 1\textwidth]{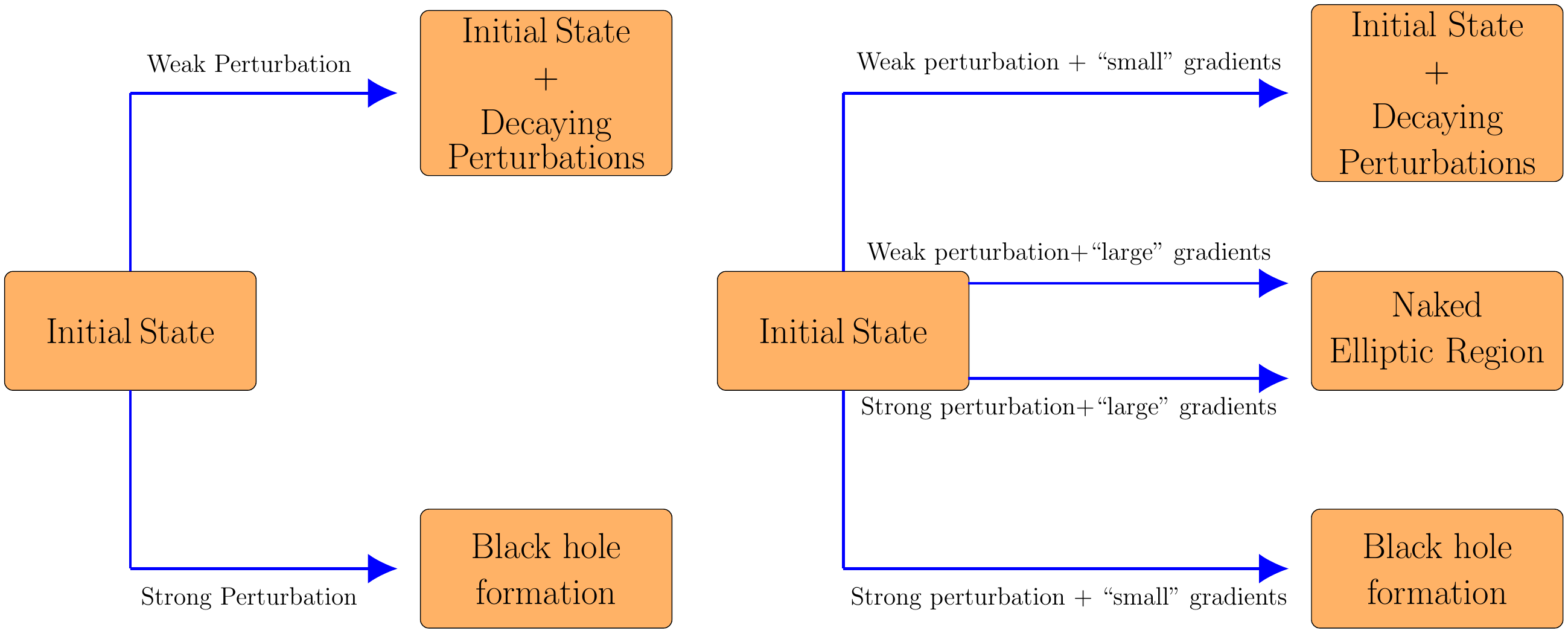}
    \caption{Flow chart illustrating the possible end states of gravitational collapse of non-BH initial data in GR (left) versus ESGB gravity (right). Gravitationally weak initial data in GR might not be weak data in ESGB gravity if the coupling constant $\ell$ is too large compared to the gradients present in initial data.}
    \label{fig:GRvsESGB}
\end{figure*}
\subsection{Existence line for static BH solutions}\label{sec:kanti-bound}
Black holes in ESGB theories are well-known to have a minimum size, given a value of $\ell$ \cite{1996-Kanti,Sotiriou:2013qea}.
To find the existence line for static BH solutions, one starts by assuming that the spacetime has a BH, with an event horizon located at $r=r_{H}$ and then one expands the field equations in a Taylor series around $r = r_{H}$. 
Let $\phi_{H}$ be the value of the Gauss-Bonnet scalar field at the event horizon and $\phi'_{H}$ denote the radial derivative of the scalar field at that same location.
Solving the field equations, one finds
\begin{equation}
    \phi'_H = \frac{-r_H^2 + \sqrt{r_H^4 - 192\,\ell^4 \,f'(\phi_H)^2}}{8\,r_Hf'(\phi_H)} \,.
\end{equation}
For the derivative of the scalar field to remain real, then $r_H$ must satisfy
\begin{equation}\label{eq:Kanti-bound}
    r_H > \left[192 \,f'(\phi_H) \right]^{1/4}\,\ell\,.
\end{equation}
Static BH solutions of ESGB gravity possess  a curvature singularity where $\phi$
blows up.
As one saturates the above bound, the curvature singularity moves closer to the event horizon 
but exactly what happens at the bound is not well understood
\cite{1996-Kanti,Sotiriou:2013qea,Julie:2022huo}. 
We also note that the value $\phi_H$ is not an independent value, as the outer boundary condition on
$\phi$ at spatial infinity affects the value of $\phi_H$ ~\cite{1996-Kanti,Sotiriou:2013qea,HegadeKR:2022xij}.
Heuristically, one expects the gradient expansion to break down as the size of the BHs becomes small compared to the coupling constant $\ell$.
This means that one naturally expects BHs slightly above the existence line to be unstable to dynamical evolution.
\section{Numerical Collapse Evolutions in ESGB gravity}
In this section, we present the results from our numerical simulations using the initial data discussed in Table.~\ref{tab:IC} and Sec.~\ref{sec:ID}. In Sec.~\ref{sec:shift-symm} we present the results for the shift-symmetric theory and then we discuss our results for the Gaussian coupling function in Sec.~\ref{sec:Gaussian}.

Before proceeding to our numerical results, we schematically explain what we are after. Figure~\ref{fig:GRvsESGB} shows a  cartoon that describes the end states of gravitational collapse in GR (left panel) and in ESGB gravity (right panel).
In GR, a sufficiently small perturbation of a stable initial state (such as flat spacetime, or a BH spacetime) results in an end state that is a weakly perturbed initial state. A sufficiently strong perturbation, however, can trigger gravitational collapse and a BH end state, even if the initial state did not contain a BH.
In ESGB gravity, on the other hand, the situation is drastically different because of the existence of an additional length scale through the coupling constant $\ell$. For sufficiently small $\ell$ (as compared to gradients of the perturbations of the initial state), the evolution is similar to the GR case described above. But for sufficiently large $\ell$ (relative to gradients of the perturbations of the initial state), both strong and weak perturbations can result in the formation of NERs.


With this schematic cartoon in mind, we can now classify the late-time evolution of some given initial data with a ``phase-space'' portrait. For concreteness, we characterize the ``strength'' of our initial data and its initial perturbations with the total ADM mass $M_0$ of the spacetime. Given this, we will then determine the outcome of the evolution of this data for a given value of the Gauss-Bonnet coupling $\ell$. For example, consider an imploding spherical shell of scalar field (in an otherwise flat spacetime) as the initial data. In GR (when $\ell=0$), this data will completely disperse and evolve into a flat spacetime end state if the ADM mass of the scalar is small enough. If, however, the ADM mass is large enough, this same initial data will evolve into a BH. The dividing line between the flat spacetime and BH end states is given by the critical collapse solution of Choptuik~\cite{Choptuik:1992jv}. In ESGB theory (when $\ell \neq 0$), there will be two new kinds of end state: a (scalar) hairy BH or NERs~\cite{Ripley:2019,Ripley:2019-ESGB,Ripley:2022cdh}. In what follows, we will use a set of numerical evolutions to construct this phase-space portrait as a function of the ADM mass of the initial data and the Gauss-Bonnet coupling constant $\ell$ for two representative coupling functions $f\left(\phi\right)$.

\subsection{Shift symmetric theory}\label{sec:shift-symm}

We first consider results for shift-symmetric ESGB gravity, i.e.~for the coupling function
\begin{align}
    \label{eq:shift_symmetric}
    f\left(\phi\right)=\phi
    .
\end{align}
Schwarzschild BHs are not stationary solutions in this theory. Instead, BHs form scalar hair, the amount of which depends on the BH mass $M$ and the Gauss-Bonnet coupling 
$\ell$ \cite{Yunes_Stein:2011we,Sotiriou:2013qea,Sotiriou_2014}. Below, we describe the phase-space portrait for the end states of shift-symmetric ESGB theory with the initial data described in Table~\ref{tab:IC} and Sec.~\ref{sec:Numerical-Results}. 

\begin{figure}[h!]
    \centering
    \includegraphics[width = 1\columnwidth]{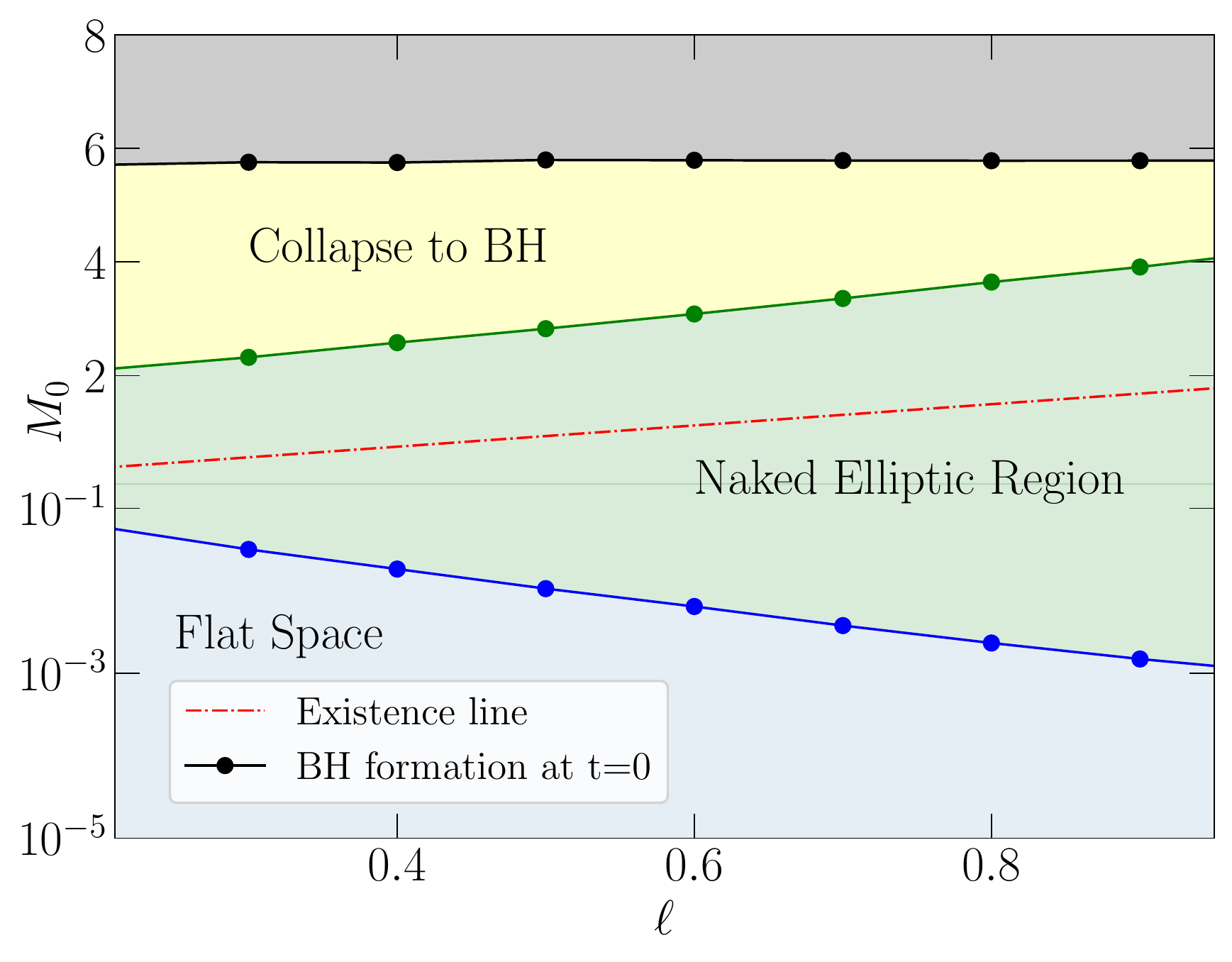}
    \caption{The phase space of gravitational collapse from GBCIC (see Table~\ref{tab:IC}) with $r_l = 8$ and $r_u = 12$ for the shift symmetric theory\footnote{We remind the reader that the length scales in all the figures are scaled by a fiducial length scale $M_{\star}$}.
    We see that the phase space consists of three possible end states:
    (1) dispersion of the scalar field to flat spacetime (light blue shaded region), 
    (2) collapse and formation of NERs, where the theory loses hyperbolicity (green shaded region),
    and 
    (3) collapse to scalarized BHs (yellow shaded region). The black shaded regions represent cases for which a BH is already present in the initial data due to the high ADM mass of the scalar field. 
    The red, dash-dotted line is given by Eq.~\eqref{eq:existence-shift-symm} and it represents the existence line for static BH solutions.
    Note that the bottom half of the figure is plotted in a log scale.
    Observe that although static BHs exist above the red, dash-dotted line, such BHs do not result from scalar field collapse for a range of $M_0$ and all $\ell$, due to the emergence of NERs.   
    }
    \label{fig:phase-space-shift-symm-GBCIC}
\end{figure}

\begin{figure}[h!]
    \centering
    \includegraphics[width = 1\columnwidth]{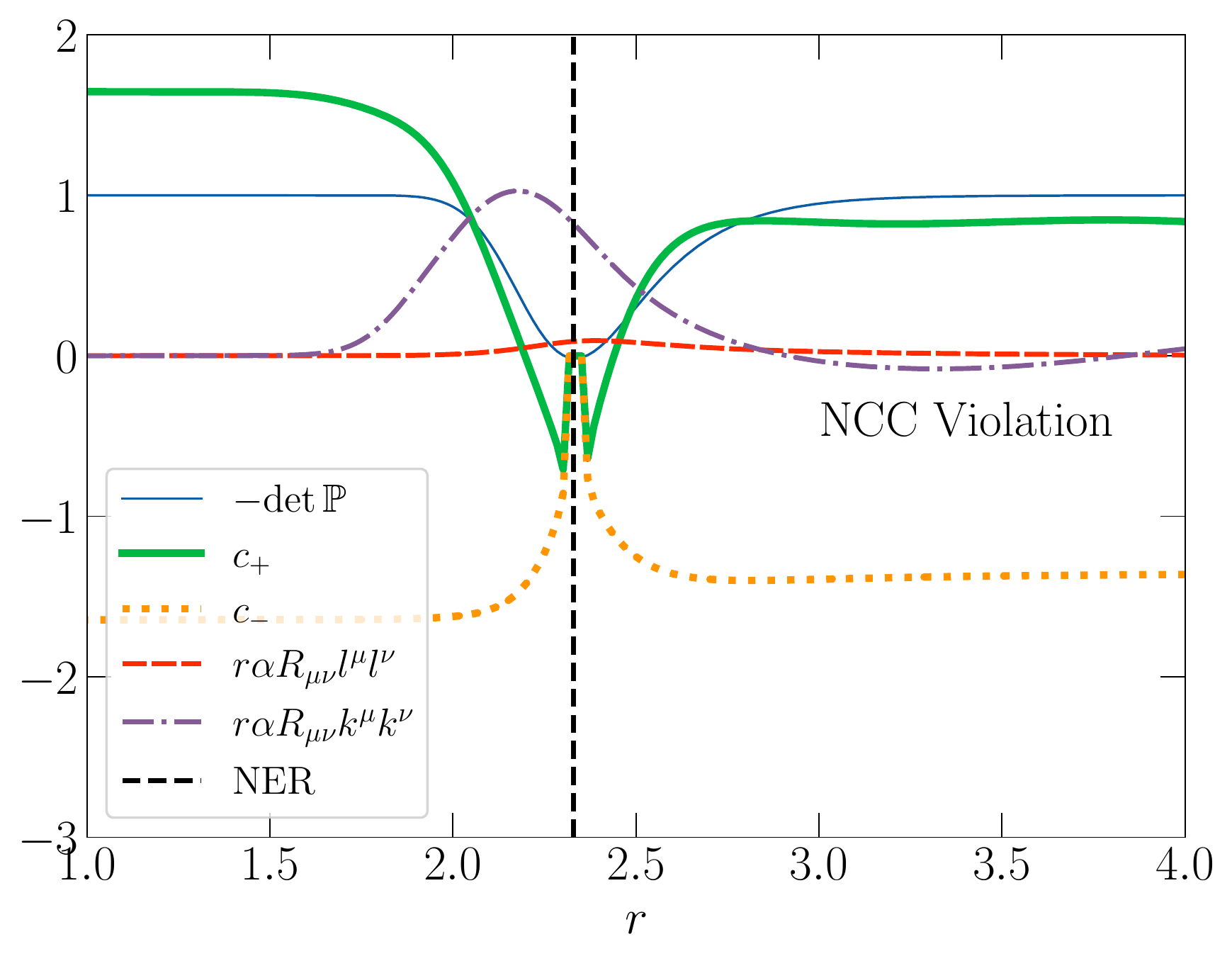}
    \caption{
    Different diagnostics for the emergence of NERs and the breakdown of hyperbolicity with GBICIC for the shift-symmetric 
    theory with $\ell = 1$ and $A = 0.06$. 
    Observe that the NCC is violated at around $r=3$ 
    while a NER forms at $r=2.3$ (shown as a dotted black line). 
    We also plot the values of $-\text{det}\, \mathbb{P}$~\eqref{eq:detP-null-frame} 
    and the values of the outgoing and ingoing characteristic speeds $c_{\pm}$~\eqref{eq:cpm}. 
    As we see from the figure $\text{det}\, \mathbb{P}$ 
    goes to zero at $r=2.3$ signalling the breakdown of hyperbolicity. }
    \label{fig:nec-non-violation}
\end{figure}

\begin{figure*}[t!]
    \centering
    \includegraphics[width = 1\columnwidth]{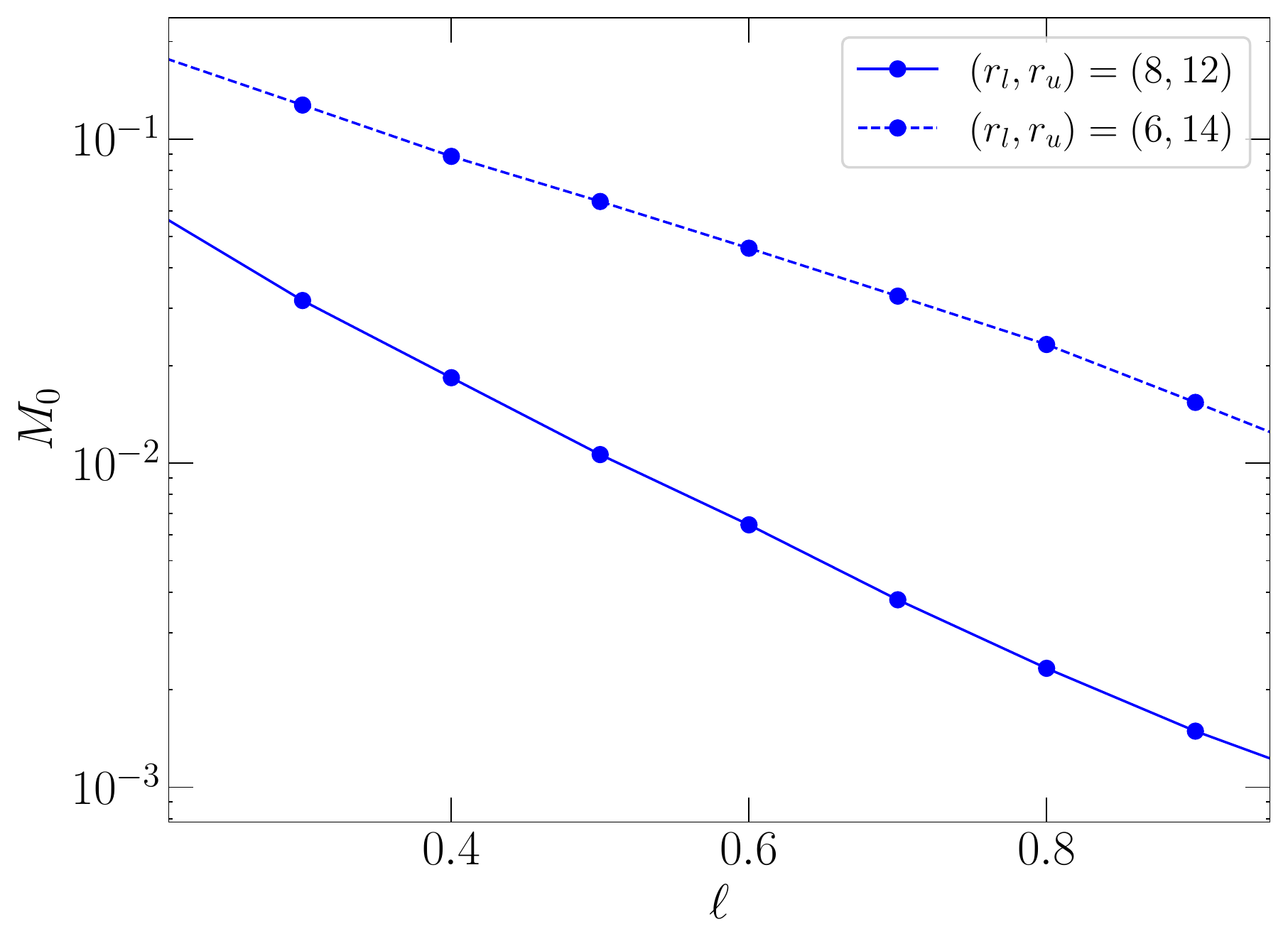}
    \includegraphics[width = 1\columnwidth]{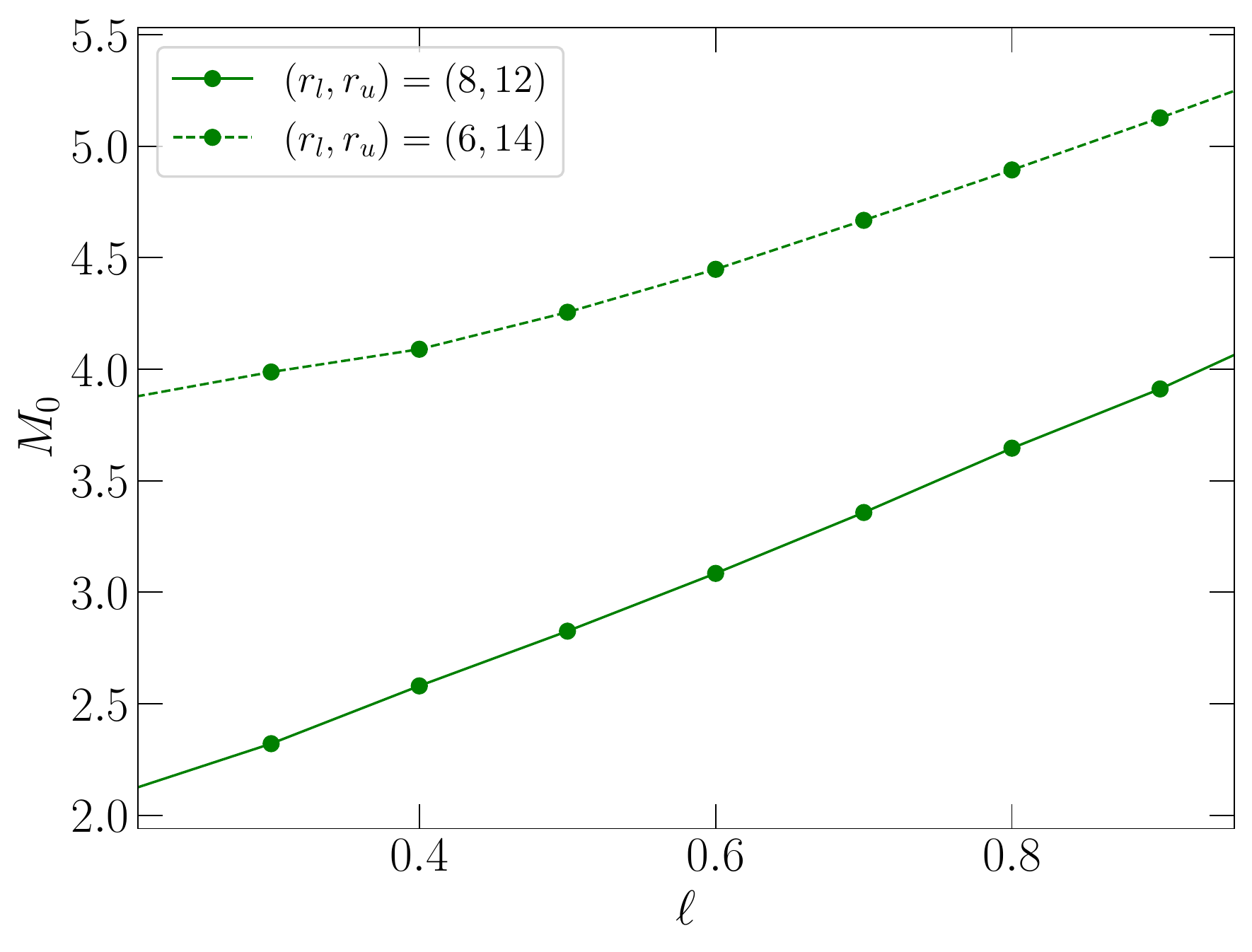}
    \caption{Shift in dividing line between the 
    flat spacetime end state and the NER end state (left) 
    and between the NER end state and the BH end state 
    (right) due to an increase in the width of the initial data.
    The bold line in both panels are the same as those shown in  Fig.~\eqref{fig:phase-space-shift-symm-GBCIC}. Observe that as the initial scalar profile is made wider (dashed curves), the dividing lines shift upwards.}
    \label{fig:comparision}
\end{figure*}
\subsubsection{Phase-Space Portrait for GBCIC in Shift-Symmetric ESGB Gravity}

We first consider the case in which a Gauss-Bonnet scalar field collapses into itself in an otherwise flat space time (the GBIC case with parameters $r_l = 8$ and $r_u = 12$, see Table~\ref{tab:IC}). Figure~\ref{fig:phase-space-shift-symm-GBCIC} shows the phase space of end states. The vertical axis is the initial ADM mass of the spacetime $M_0$, and the horizontal axis is the value of the Gauss-Bonnet coupling constant $\ell$.
For a given value of the coupling constant $\ell$ and sufficiently weak initial data (i.e.~small $M_0$), the evolution ends in the complete dispersal of scalar waves and the end state is Minkowski spacetime (light blue shaded region). For very small values of $\ell$ and moderate $M_0$, one approaches the critical Choptuik solution, whose numerical characterization would require the evolution of initial data with adaptive mesh refinement (AMR). Since our numerical implementation does not include AMR, we focus on cases with $\ell > 0.2$ to ensure numerical convergence.  As the strength of the initial data (characterized by $M_0$) increases, the collapse leads to the formation of NERs, and the theory loses hyperbolicity (light green shaded region). 

We determine the existence of NERs through various diagnostics, shown in Fig.~\ref{fig:nec-non-violation} for one example run. 
In particular, the figure presents the determinant of principal symbol $\det{\mathbb{P}}$ discussed in Eq.~\eqref{eq:detP-null-frame} (blue line), the outgoing characteristic speed $c_+$ (bold green line), the ingoing  characteristic speed $c_-$ (dashed orange line), the ingoing (dashed red line) and outgoing (dashed purple line) null convergence conditions, and the radial coordinate of the NER region (vertical black line) on a time slice when the NER first appears.
As we discussed in Sec.~\ref{sec:null-frame-analysis},
the breakdown of hyperbolicity is related to strong focusing of null-geodesics, at least in spherical symmetry. 
This is visible in the figure, as we see that $\det\mathbb{P}$ is zero around the peak of the outgoing convergence condition.
Observe, in particular, that the emergence of NERs 
does not occur when the NCC is violated.

Before we describe Fig.~\ref{fig:phase-space-shift-symm-GBCIC} further, 
we recall the existence condition for the shift symmetric theory that we presented in Eq.~\eqref{eq:Kanti-bound}, which reduces to
\begin{equation}
    r_H >(194)^{1/4} \ell\sim 3.73\, \ell\,,
\end{equation}
approximating $r_H \sim 2 M_{\text{BH}}$. This inequality can then be rewritten to find approximately 
\begin{equation}\label{eq:existence-shift-symm}
    M_{\text{BH}} > 1.87\, \ell\,.
\end{equation}
This approximate existence line for static BH solutions tells us that BHs below a certain mass cannot exist in the shift symmetric theory if one treats the equations of motion as exact. Equation~\eqref{eq:existence-shift-symm} is shown in Fig.~\ref{fig:phase-space-shift-symm-GBCIC} as a dash-dotted red line.

We now continue discussing Fig.~\ref{fig:phase-space-shift-symm-GBCIC}. 
For sufficiently strong initial data (i.e.~for sufficiently large $M_0$, given a fixed $\ell$), the evolution leads to the formation of a trapped 
surface, which ``hides'' the elliptic region. The end state in this case is the formation of a scalarized BH (shaded yellow region)\footnote{Since, our initial data is ingoing, we find the initial mass $M_0$ is close to the final BH mass $M_{\text{BH}}$.}. The dividing line (shown in green) between the formation of NER and the formation of a stable scalarized BH lies \textit{above} the existence line of Eq.~\eqref{eq:existence-shift-symm}, shown as a dash-dotted red line in the figure. This means that although \textit{static} scalarized BH exist, dynamical collapse does not allow for their formation. As we continue to increase the strength of the initial data, then the data itself already contains a BH, so the evolution proceeds through the absorption of the scalar field and a BH end state (black shaded region). 

The general conclusions presented above are robust to the details associated with the initial data, but the precise location of the dividing lines between end states in Fig.~\ref{fig:phase-space-shift-symm-GBCIC} is not. We compare how the diving lines shift by increasing the width of the initial profile (i.e.~changing $r_l$ and $r_u$) in Fig.~\ref{fig:comparision}. When we increase the width of the scalar field, the latter becomes initially less focused, so to obtain BH formation, we must endow the field with a larger ADM mass. The dividing line between the flat spacetime end state and the NER end state shifts (left panel) and between the NER end state and the BH end state (right panel) shifts upwards when the initial pulse is wider. Therefore, the regime inside which static BHs exist but scalar field collapse leads to NERs becomes larger, and the conclusions presented above remain unchanged. 
\begin{figure}[h!]
    \centering
    \includegraphics[width = 1\columnwidth]{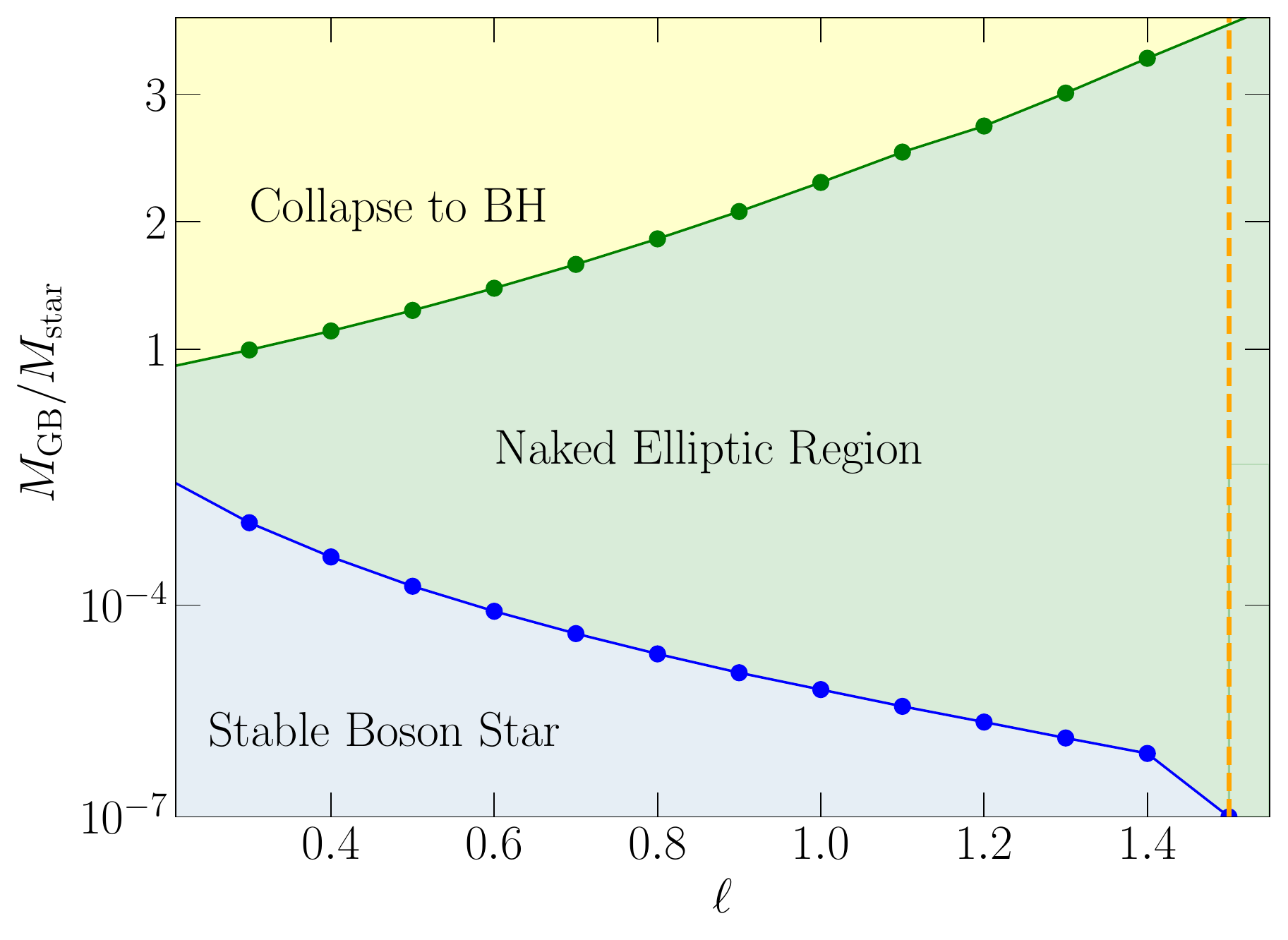}
    \caption{Phase space portrait of possible end states in shift symmetric theory when perturbing a stable boson star with an infalling Gauss-Bonnet scalar field. Observe that, for sufficiently large $\ell$, extremely small values of initial Gauss-Bonnet scalar field amplitude can evolve into NERs in the interior of the star. The orange dashed line indicates the maximum value of $\ell$ above which all perturbations
    from flat space by the Gauss-Bonnet scalar that we studied end in NER formation.  }
    \label{fig:bs-ner-shift-symm}
\end{figure}

\subsubsection{Phase-Space Portrait for SBSCIC in Shift-Symmetric ESGB Gravity}

We now consider how the results obtained above change if considers the collapse of the GB scalar field onto an (otherwise stable), self-gravitating object, such as a star (the SBSCIC case). As a toy model for a star, we consider a boson star because of their relatively simpler equations of motion, as compared to those of a relativistic fluid.
The boson star profiles we consider are completely determined by the central
value of the scalar field $\rho_c := \rho(0)$ 
and the mass $m_b$, as we briefly review in Appendix~\ref{appendix:Boson-star}.
We set $\rho_c = 0.3$ and $m_b = 0.5$ which roughly translates to an initial ADM mass of $M_{\text{star}} = 1.24$ and $R_{\text{star}} = 16.8$.
We remind the reader that all quantities with the dimensions on length are scaled with a fiducial mass $M_{\star}$ as mentioned in the introduction.
We now ask how perturbing this star with an infalling Gauss-Bonnet scalar field affects the boson star.
We setup SBSCIC initial data with $r_l = 18$ and $r_u = 22$ and vary the amplitude to see the transition between relaxation back to a boson star and evolution to NERs or BH formation. 

The phase space portrait of end states for perturbed boson stars is presented in Fig.~\ref{fig:bs-ner-shift-symm}. Observe that this figure is qualitatively similar to Fig.~\ref{fig:phase-space-shift-symm-GBCIC}. For a fixed and small value of $\ell$, weak perturbations (with small $M_{\rm GB}/M_{\rm star}$) lead back to a boson star end state. But as the strength of the perturbation is increased, the perturbed boson star evolves into a NER. 
Eventually, for sufficiently strong perturbations, the boson star collapses to a BH. An interesting feature of this type of initial data that is not found in the flat spacetime case is that, as the strength of the Gauss-Bonnet coupling $\ell$ is increased, the size of the phase space in which the end state is a stable boson star rapidly decreases. Eventually, once $\ell_{\text{max}} > 1.45$ all perturbations with $M_{\rm GB}/M_{\rm star} \geq 10^{-6}$ evolve into NERs. 

Such a shrinkage of parameter space is not present when perturbing flat spacetime. This is because setting the scalar field amplitude to zero just returns flat spacetime as the solution. Setting the scalar field amplitude to zero in the boson star case should return a boson star, but the latter generically has a nonzero spacetime Gauss-Bonnet curvature, which  sources the growth of the scalar field. This leads us to conjecture that such a breakdown may happen for any sufficiently compact object in the shift symmetric theory, given a large enough value of the coupling constant $\ell$. The value of $\ell_{\max}$ for hyperbolic evolution will depend on the strength of the  background curvature for a given scalar field initial data. 
\subsubsection{Phase-Space Portrait for BHIC in Shift-Symmetric ESGB Gravity}
\begin{figure}
    \centering
    \includegraphics[width = 1\columnwidth]{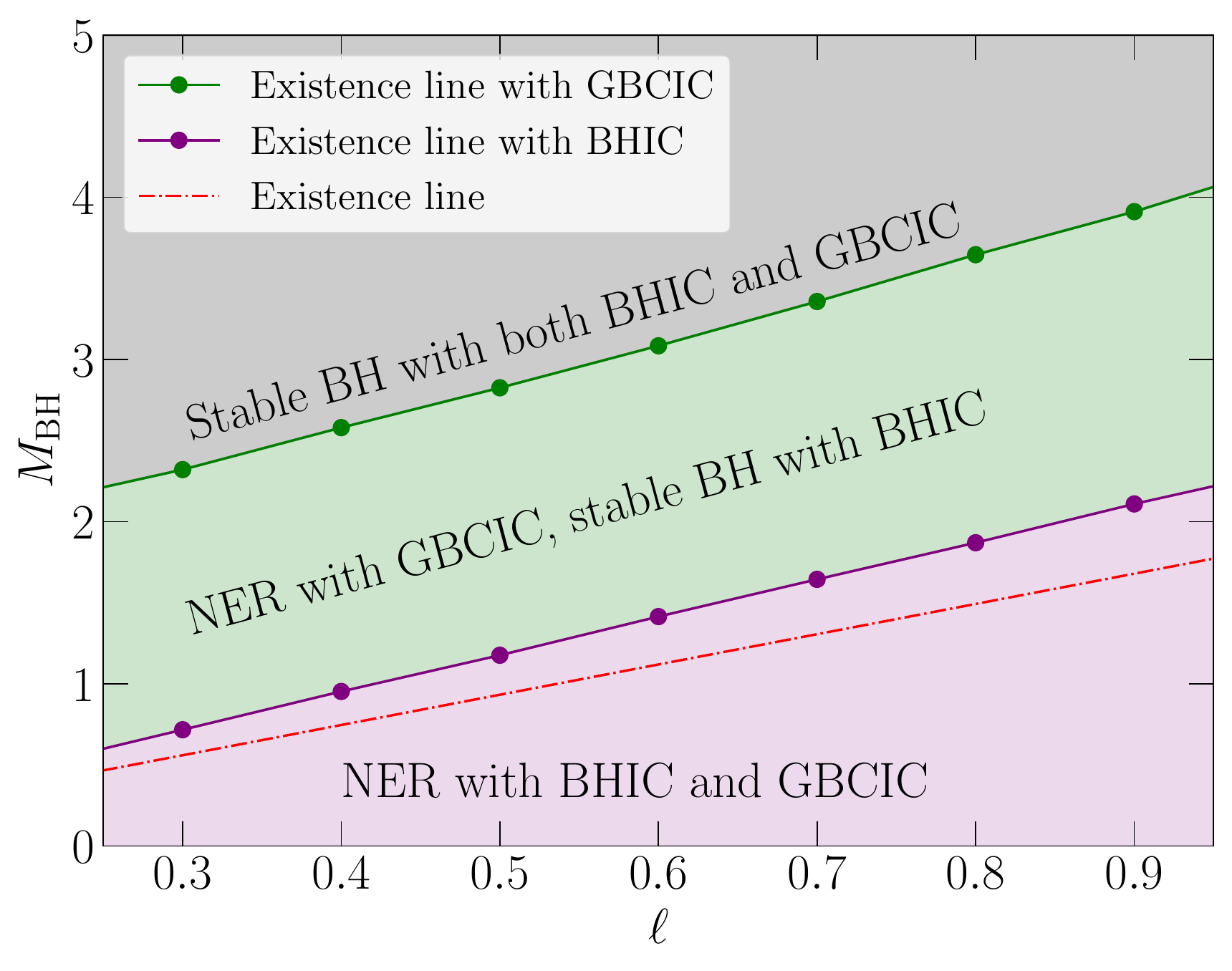}
    \caption{The smallest possible BHs allowed in shift symmetric theory 
    as a function of the coupling constant with BHIC. 
    We set the parameters $r_l = 8$ and $r_u = 12$ and $A = 10^{-3}$. 
    Dynamical evolution from BHIC shows that theory loses hyperbolicity below the purple line. 
    We also show the existence line for static BH solutions (red dash-dotted line) 
    and the existence line from GBCIC (shown in green, same as in Fig.~\ref{fig:phase-space-shift-symm-GBCIC}). 
    The purple line depends on the strength of the initial scalar field profile 
    and can get closer to the green line if one increases the initial amplitude 
    of the scalar field.  }
    \label{fig:bh-ner-shift-symm}
\end{figure}

We now investigate how the transition to NERs occurs when the Gauss-Bonnet scalar field falls into a hairy BH (the BHIC case). Since BHs of arbitrary size cannot exist in shift symmetric theory (see e.g.~Eq.~\eqref{eq:existence-shift-symm}), we now construct a phase portrait with the vertical axis representing the BH mass, while fixing the amplitude of the scalar field perturbation. More concretely, in all our numerical experiments, we set $r_l = 8$ and $r_u = 12$, and we fix the initial amplitude of the perturbation to $10^{-3}$. 

The phase space portrait in the BHIC case is presented in Fig.~\ref{fig:bh-ner-shift-symm}. For sufficiently large BHs, the scalar field perturbation does little, and the end state is again a BH (black and green shaded regions). This occurs even in a regime of parameter space in which the collapse of a scalar field in an otherwise flat spacetime would have led to the formation of NERs (green shaded region). As the mass of the BH is decreased, however, NERs arise generically. The smallest BHs that are stable to the scalar field perturbation without forming NERs can be approximately fitted by the line
\begin{equation}
    M_{\text{BH}} = (2.35 \pm 0.008) \,\ell\,,
\end{equation}
which is represented by a purple line in Fig.~\ref{fig:bh-ner-shift-symm}. Observe that this line lies above the existence line for static solutions of Eq.~\eqref{eq:existence-shift-symm} (red dot-dashed line), but \textit{below} the existence line of the GBCIC case (green line, shown also in green in Fig.~\ref{fig:phase-space-shift-symm-GBCIC}). This result is generic, but how close this line is to the existence line of the GBCIC case depends on the strength of the initial scalar field. 
\subsection{Gaussian Theory}\label{sec:Gaussian}
We next discuss our numerical results in the Gaussian theory, which was first
introduced in \cite{Doneva:2017bvd} as
\begin{equation}\label{eq:gaussian-func}
    f(\phi,\mu) = \frac{1 - e^{-\mu \phi^2}}{2\mu}
    \,,
\end{equation}
where $\mu$ is a constant. Since $\ell$ multiplies this coupling function, the
Gaussian is then parametrized by two constants $(\ell,\mu)$, where $\ell/\mu$ controls
the size of the GR deformation, and $\mu$ controls the shape of the coupling function.  
In most of this subsection, we will set $\mu = 3$, and comment on other values of $\mu$
at the end. 

Static and spherically symmetric BH solutions for the Gaussian 
theory are of two different classes. One of them consists simply
of the Schwarzschild solution with a zero scalar field ($\phi = 0$). 
The other consists of a non-Schwarzschild BH solution with non-zero scalar
hair. The existence condition of Eq.~\eqref{eq:Kanti-bound} implies that these 
scalarized solutions occur in set of banded regions in the 
$\ell$-BH mass ($\ell$ - $M_{\text{BH}}$) 
plane (see for example, Fig. 2 of Ref.~\cite{Silva_2018}).
The first (GR) branch of solutions is actually unstable to
the growth of scalar hair under a small scalar perturbation
in some regions of parameter space, a process known as 
\emph{spontaneous BH scalarization}.

To better understand how BHs can be unstable to scalarization in this theory, 
we consider the scalar equation of motion~\cite{Silva_2018,Doneva:2017bvd}:
\begin{equation}
    \Box{\phi} + \ell^2 \phi\, e^{-\mu\phi^2} \mathcal{G} = 0\,.
\end{equation}
We rescale $\phi $ to $\phi/\sqrt{\mu}$ and expand about small $\phi$ to get
\begin{equation}
    \left(\Box + \ell^2 \, \mathcal{G}\right) {\phi} =0\,.
\end{equation}
The $\ell^2\mathcal{G}$ term can act like an ``effective mass'' in the linearized
equation, and if the mass is tachyonic ($\ell^2\mathcal{G}>0$), then
the scalar field can be unstable to growth.

As the effective mass varies in space, and due to the presence of the boundary
conditions at infinity and the BH horizon, not all BHs
are unstable to hair growth in this theory.
A detailed analysis shows that Schwarzschild BHs are unstable to small
linear scalar perturbations when 
\begin{equation}\label{eq:instability-schwarzschild}
   M_{\text{BH}} \leq 1.174\, \ell
   .
\end{equation}
In general, coupling functions that can be expanded to give a coupling
of the form $\sim \phi \mathcal{G}$ to leading order in $\phi$ can 
lead to spontaneous BH scalarization, as discussed for example \cite{Silva_2018}.
For BHs with $M_{\text{BH}} \leq 1.174\, \ell$, scalar hairy BH solutions can be found to occur in bands.
These solutions are perturbatively
stable, so one concludes that generally BHs in 
that mass range should have scalar hair~\cite{Doneva:2017bvd,Silva_2018}. 

What these earlier perturbative studies do not address, however, is whether the theory remains predictive (weakly-coupled) during the BH scalarization process.
Earlier work suggests that there is only a narrow range of masses for which the theory
remains weakly coupled and can have scalar hairy BHs \cite{East:2021bqk}.
Here, we present a more exhaustive analysis of this question, which strongly suggests that 
the phenomena of spontaneous scalarization falls very close to the breakdown 
of the gradient expansion, used to justify the truncation of ESGB gravity at quadratic
order in curvatures, splitting the analysis into the three types of initial data we
considered before (GBCIC, SBSCIC, and BHIC). 
We also note that the process of spontaneous scalarization has been questioned by deriving positivity bounds~\cite{Herrero_Valea_2022}.
\subsubsection{Phase-Space Portrait for GBCIC in Gaussian ESGB Gravity}
\begin{figure}[h!]
    \centering
    \includegraphics[width = 1\columnwidth]{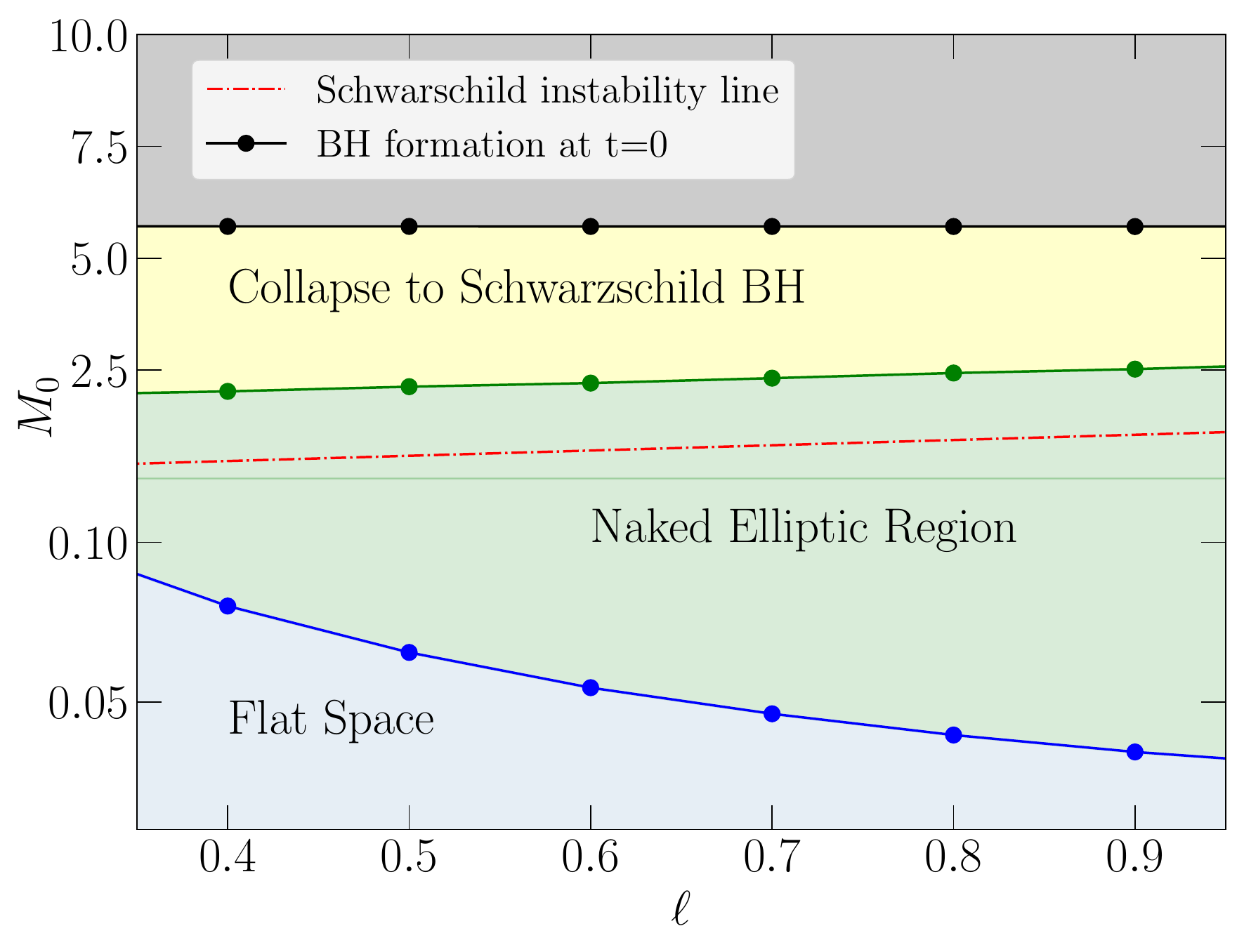}
    \caption{Phase space of scalar collapse in an otherwise flat spacetime (the GBCIC case) in Gaussian ESGB theory with $\mu = 3$.
    Observe that, as in the shift-symmetric case, the phase space contains three possible end states: dispersion to flat space (blue region), collapse leading to NERs and the breakdown of hyperbolicity (green region), and collapse to a non-hairy, \textit{Schwarzschild} BHs (yellow region). The dash-dotted red line indicates the region below which the Schwarzschild solution is unstable to scalar hair growth [see Eq.~\eqref{eq:instability-schwarzschild}]. Observe that for all simulations, scalar field collapse never leads to spontaneously scalarized BHs due to the emergence of NERs at small scalar field masses.}
    \label{fig:phase-space-gaussian-GBCIC}
\end{figure}

We begin by considering again the collapse of the Gauss-Bonnet scalar field in an otherwise flat spacetime (the GBCIC case), but this time in Gaussian ESGB theory, where again we have set the initial conditions to $r_l = 8$ and $r_u = 12$. The phase space portrait of end states that we find is shown in Fig.~\ref{fig:phase-space-gaussian-GBCIC}. As before, for weak data (small $M_0$), the scalar field disperses and the spacetime remains flat (blue region). As the strength of the data is increased, the collapse of the scalar field leads to a NER (green region). Eventually, for even larger values of $M_0$, the scalar field collapses to a BH (yellow region) or the initial data already contains a BH (black region). 

Observe that the phase space portrait of Fig.~\ref{fig:phase-space-gaussian-GBCIC} is qualitatively similar to that found in the shift-symmetric theory in Fig.~\ref{fig:phase-space-shift-symm-GBCIC}. Observe that the curve separating NER formation from dispersion to flat spacetime is slightly higher in Gaussian ESGB theory than in the shift-symmetric theory. This is because, for weak initial data, the Gaussian coupling function exponentially suppresses the Gauss-Bonnet corrections to the equations of motion (i.e.~when $\phi$ grows large, $e^{-3\phi^2} \ll 1$). 

We emphasize that we \textit{do not} form any scalarized BHs as end states of gravitational collapse in our simulations with GBCIC. As we see in the figure, the curve separating collapse to BHs lies much above the line below which the Schwarzschild solution is unstable, as given in Eq.~\eqref{eq:instability-schwarzschild} (red dash-dotted line). The precise location of this curve, however, depends on the details of the initial scalar field profile. As we will show in Sec.~\ref{sec:BHIC-Gaussian}, there are finely-tuned choices of scalar field initial data that do lead to the formation of scalarized BHs. 

\subsubsection{Phase-Space Portrait for SBSCIC in Gaussian ESGB Gravity}
\begin{figure}[h!]
    \centering
    \includegraphics[width = 1\columnwidth]{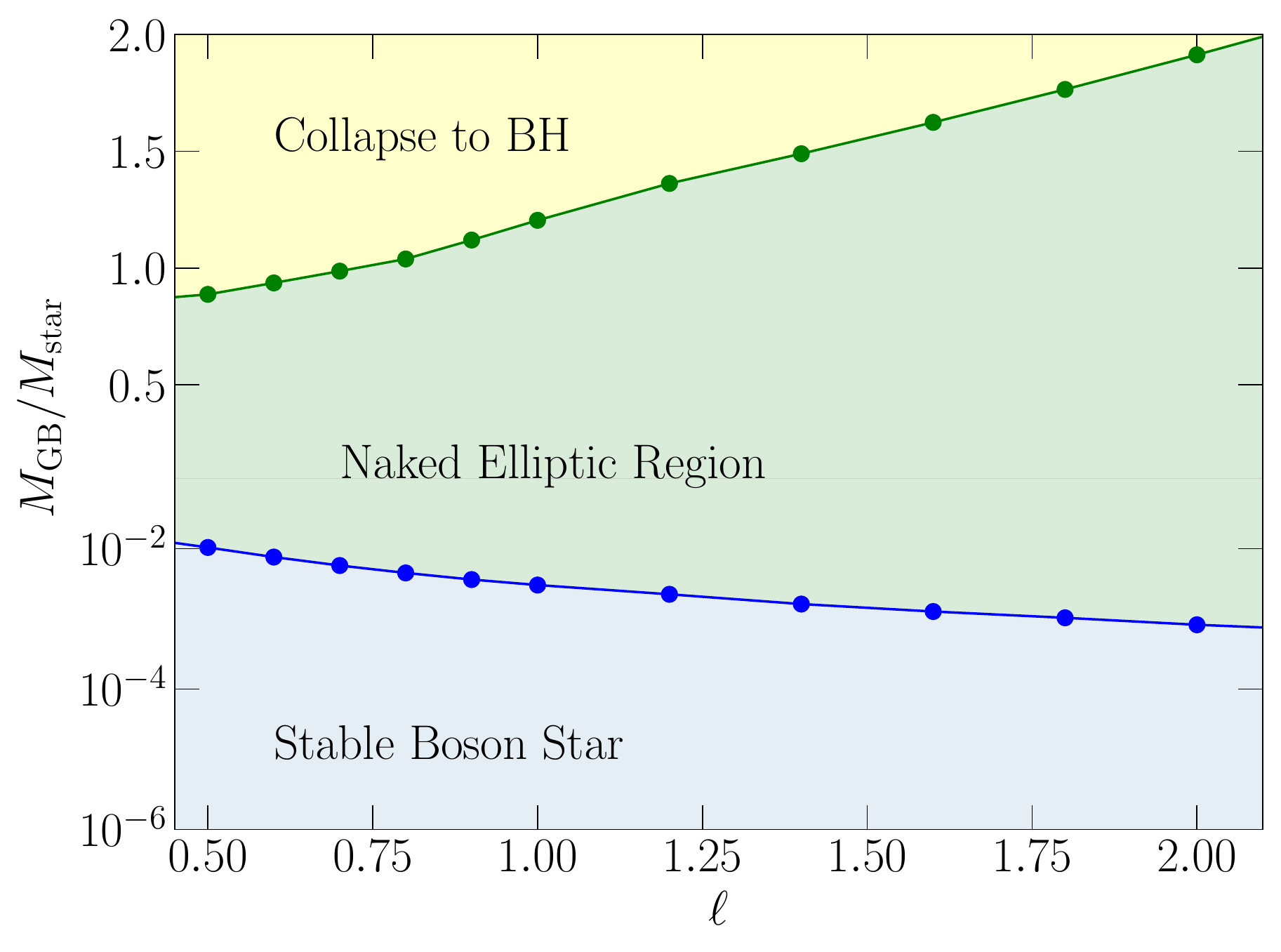}
    \caption{Phase space portrait of end states for a scalar field perturbation of a boson star (SBSCIC case) in Gaussian ESGB theory, with initial scalar field parameters $r_l = 18$, $r_u = 22$ and $A = 10^{-3}$. The strength of the scalar field amplitude required to destabilize the star decreases as one increases the coupling constant $\ell$. The dividing line between evolution into NER and dispersion back to the stable boson star is shown in blue. Observe that for all values of $\ell$, NERs generically arise for sufficiently strong initial scalar field perturbations.}
    \label{fig:bs-ner-gaussian}
\end{figure}

We now consider again the collapse of the Gauss-Bonnet scalar field into an otherwise stable boson star (the SBSCIC case), but this time in Gaussian ESGB theory. The phase space portrait we obtain is shown in Fig.~\ref{fig:bs-ner-gaussian}. As in the shift-symmetric case, for sufficiently weak initial data, the scalar field perturbation disperses and the end state is a boson star (blue region). For any fixed value of $\ell \neq 0$, however, as the strength of the initial perturbation increases, the evolution forms NERs inside which hyperbolicity is lost (green regions). Unlike in the shift symmetric case, however, we do not find a maximum value of $\ell$ for which NERs always appear (i.e.~there is no analog of the maximum-$\ell$, vertical line of Fig.~\ref{fig:bs-ner-shift-symm}). Eventually, for sufficiently strong initial perturbations, the scalar field collapses to a BH (yellow region) and all NERs are hidden inside the horizon. 



\subsubsection{Phase-Space Portrait for BHIC in Gaussian ESGB Gravity}\label{sec:BHIC-Gaussian}
\begin{figure}[h!]
    \centering
    \includegraphics[width = 1\columnwidth]{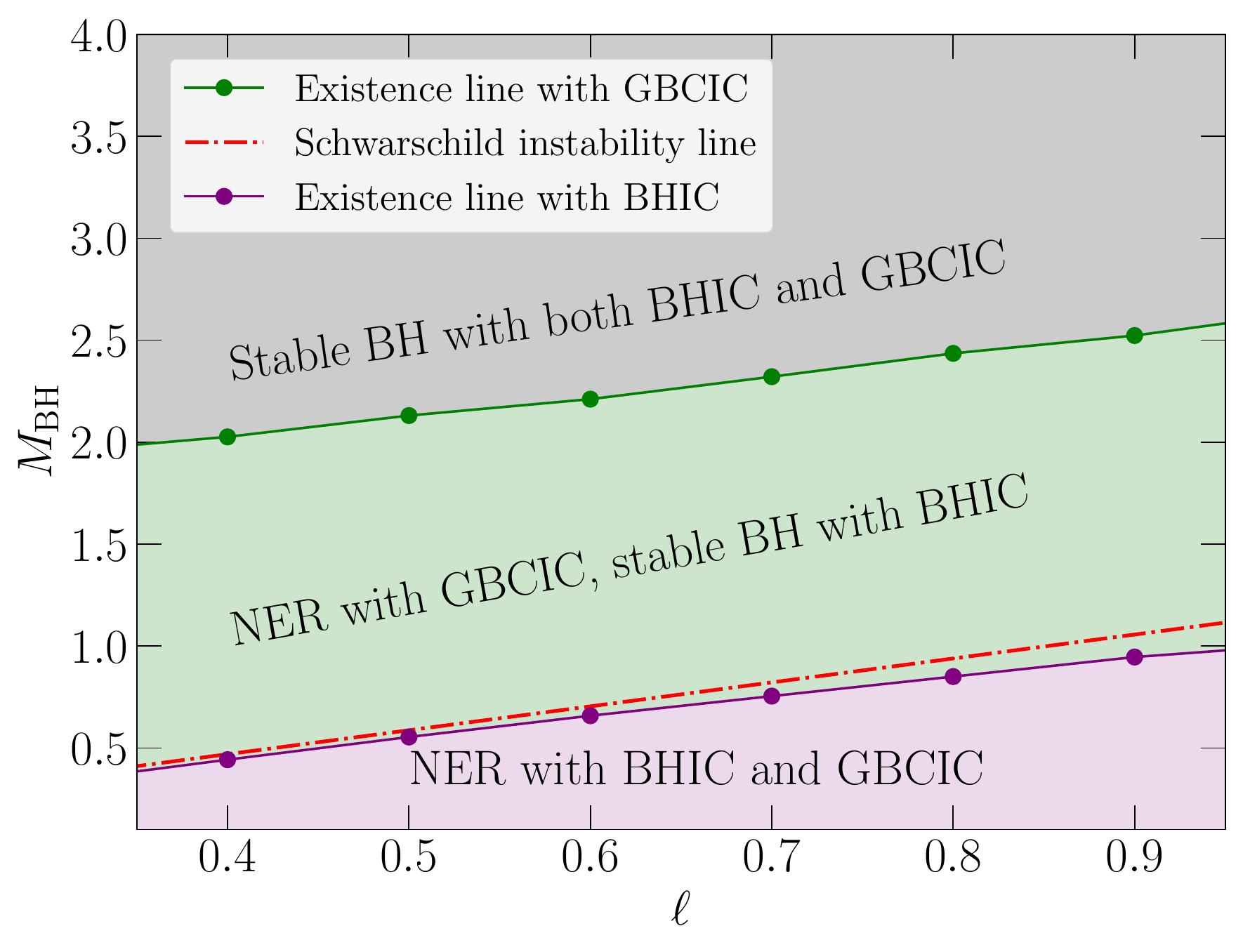}
    \caption{Stability of small BHs is Gaussian theory with $\mu = 3$ with BHIC. 
    We set a fixed scalar field perturbation with $r_l = 8$, $r_u = 12$ and $A = 10^{-3}$. 
    The subsequent evolution of the scalar field results in NER for 
    BH masses below the purple line. We also show the Schwarzschild instability
    line~\eqref{eq:instability-schwarzschild} as a dashed red line. 
    The green line separating collapse to NER with GBCIC is shown in green 
    (same line is shown in Fig.~\ref{fig:CIC-gaussian-48}. 
    The purple line can be shifted closer to the green line by increase the amplitude 
    of the scalar field perturbation.}
    \label{fig:bh-ner-gaussian}
\end{figure}

We finally consider again an infalling scalar field perturbation into an otherwise stable BH (the BHIC case), but this time in Gaussian ESGB theory. The phase space portrait is shown in Fig.~\ref{fig:bh-ner-gaussian}. As noted before, the existence line from BHIC (shown in purple) depends on the strength of initial data and can lie anywhere between the purple line and the green line. Nevertheless, for the present choice of parameters ($A = 10^{-3}$, $r_l = 8$ and $r_u = 12$), we find that one can form some scalarized BHs for which the elliptic regions are not naked, but instead are hidden inside the AH of the scalarized BHs. This set of scalarized solutions lie very close to the Schwarzschild instability line, and their mass can be best fit by the line
\begin{equation}
    M_{\text{BH}} = \left(1.08 \pm 0.01 \right)\, \ell\,.
\end{equation}
These results are consistent with those of~\cite{East:2021bqk}, which found that the evolution of spherically symmetric BHs smaller than $M_{\text{BH}} = 1.09\, \ell$ leads to NERs. 

Our results suggest that one has to be careful in interpreting the results obtained in the decoupling limit, such as in Ref.~\cite{Silva:2020omi,Elley:2022ept}. For the model we consider in this paper, the decoupling limit would result in exponential growth of scalar field on the GR background, if the mass of the GR solution is below the Schwarzschild instability line of Eq.~\eqref{eq:instability-schwarzschild}. This, however, does not mean that the BH spontaneously scalarizes if one back reacts the scalar field, because NERs may in fact appear. In reality, what the decoupling analysis implies is that the solution either spontaneously scalarizes, or the theory exits its domain of validity due to the loss of hyperbolicity.
\subsubsection{Generalization to other values of $\mu$}

At this junction, one may wonder how our results and conclusions would change if we changed the value of $\mu$ in Eq.~\eqref{eq:gaussian-func}. To understand this, we look at the Lagrangian for a general $\mu$:
\begin{equation}
    L =  R - \left(\nabla \phi\right)^2 + \frac{\ell^2}{\mu} \left(1-e^{-\mu \phi^2}\right)\mathcal{G}
\end{equation}
The scalar field equations of motion is then given by
\begin{equation}
    \Box{\phi} + \ell^2 \phi\, e^{-\mu\phi^2} \mathcal{G} = 0\,. 
\end{equation}
Rescaling the scalar field via $\Tilde{\phi} = \sqrt{\mu/3}\, \phi $, we then find
\begin{equation}
    \Box{\Tilde{\phi}} + \ell^2 \Tilde{\phi}\, e^{-3 \Tilde{\phi}^2} \mathcal{G} = 0\,. 
\end{equation}
Therefore, the $\mu$ constant in reality can be re-absorbed through a field redefinition, and only the coupling $\ell$ determines the evolution equations. As a corollary, if one were to repeat the linear stability analysis of~\cite{Silva_2018,Doneva:2017bvd} with these equations, one would  indeed find that Schwarzschild BHs suffer and tachyonic instability if 
\begin{equation}
    M_{\text{BH}} \leq 1.174 \ell\,.
\end{equation}

\begin{figure}
    \centering
    \includegraphics[width = \columnwidth]{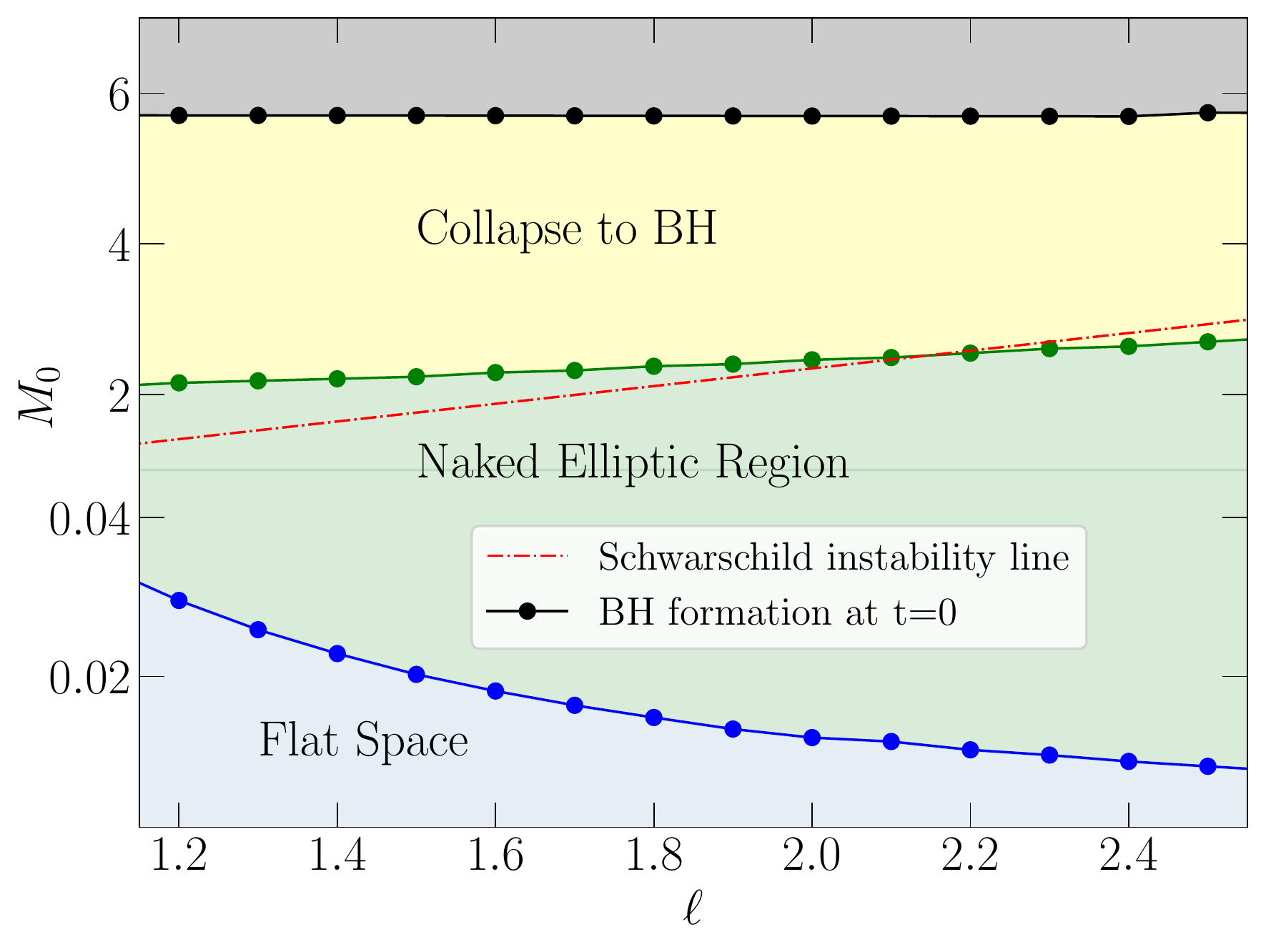}
    \caption{Phase space portrait for the collapse of the Gauss-Bonnet scalar field on an otherwise flat spacetime (the GBCIC case) in the Gaussian theory with $\mu = 48$. Observe that the Schwarzschild instability line crosses the line dividing NER formation at $\ell \sim 2.1$. Therefore, for this value of $\mu$ one can collapse the Gauss-Bonnet scalar field into scalarized BHs. The magnitude of the scalar hair, however, is suppressed by a factor of $\mu^{-1/2}$, thus reducing the impact of the Gauss-Bonnet correction on observable effects.}
    \label{fig:CIC-gaussian-48}
\end{figure}

Although the onset of the BH scalarization instability is unchanged by the value of $\mu$, the \emph{amplitude} of the scalar field around a scalarized BH does change, thus reducing the effect of the scalar Gauss-Bonnet corrections to the equations of motion. This fact was used in~\cite{Kuan_2021} to evolve a collapsing fluid simulation through the formation of a BH in the Gaussian theory. By picking large values of $\mu$, the authors were able to evolve the full theory without any loss of hyperbolicity. We show the phase space for $\mu = 48$ in the GBCIC case in Fig.~\ref{fig:CIC-gaussian-48}. Observe that the Schwarzschild instability line over takes the line dividing NERs from BH collapse. Therefore, one can form scalarized BHs with $\mu = 48$ without the loss of hyperbolicity for sufficiently large $\ell$. The amount (amplitude) of the scalar hair on the BH, however, decreases by a factor of $\sqrt{1/\mu}$ as compared to the $\mu=3$ theory, which leads to smaller observable effects.
\section{Conclusions and Future Directions}\label{sec:Conclusions}

ESGB gravity captures the leading-order,
scalar-tensor interactions in low-energy theories of quantum gravity
\cite{Zwiebach:1985uq,Gross:1986mw,Weinberg_2008,Cano:2021rey}.
While the equations of motion for the theory can be solved 
for weakly-coupled solutions
(up to truncation error)
using the techniques of numerical relativity
\cite{Kovacs_2020_PRL,Kovacs:2020ywu,East:2021bqk,CCZ4-2022,Ripley:2022cdh}, 
they can break down when the curvature scales grow too large 
\cite{Ripley:2019,Ripley:2019-ESGB,Ripley:2019aqj,East:2021bqk}.
This breakdown can be interpreted as indicating that the theory has entered a 
\emph{strong coupling regime}, 
where higher-order corrections to the equation of motion become important.

In this paper, we performed a detailed investigation of the breakdown of the equations
of motion for two ESGB theories (a shift-symmetric and a Gaussian one), 
and derived a gauge-invariant expression for the
principal symbol for general ESGB theories of gravity in spherically-symmetric spacetimes.
Our expression for the principal symbol show that (at least in spherical symmetry),
the breakdown of the equations of motion arises first in regions where the geodesic focusing is strong compared to the length scales set by the gradient present in the field (see Eqs.~\eqref{eq:detP-null-frame} and \eqref{eq:RkkRll-v2} for a precise condition).
Moreover, the breakdown is not tied to the violation of the Null Convergence Condition
which underlies the BH area theorem \cite{Hawking:1971tu}, 
and which is generically violated in solutions to ESGB theories of gravity~\cite{Alexander:2021ssr}. 
While the formulas we derived (Eqs.~\eqref{eq:detP-null-frame} and \eqref{eq:RkkRll-v2})
for the breakdown of the theory only hold for spherically-symmetric spacetimes, 
it may serve as a useful heuristic diagnostic to determine the
causes of breakdown in full $3+1$ simulations of the theory.

Moreover, in this paper, we also performed a detailed study of the non-linear, dynamical 
stability of compact objects. In particular, we considered three types of spherically-symmetric
initial data (see Table~\ref{tab:IC} and Fig.~\ref{fig:cartoon-collapse}): the in-fall of a shell of Gauss-Bonnet scalar on an otherwise flat spacetime, 
on an otherwise stable boson star, and on an otherwise stable, hairy black hole. We then 
numerically simulated the 1+1 evolution of the ESGB field equations of this data 
(in both a shift-symmetric and Gaussian theory) to determine whether the end state was
the same as the unperturbed initial data (ie. that without the infalling scalar field), whether
it was a black hole (identified through the formation of an apparent horizon), 
or whether it was the formation of a NER, which we determined through the
gauge invariant approach mentioned above. 


When considering the collapse of the Gauss-Bonnet scalar in an otherwise flat spacetime,  
we found that there is a ``gap'' in the phase space between gravitational collapse to BHs 
and dispersion to flat spacetime, extending earlier work \cite{Ripley:2019,Ripley:2019-ESGB}.
We also showed that the size of this gap increases as the value of the coupling constant
increases. Our analysis also found that the hyperbolicity breaks down 
precisely in the region where spontaneous scalarization is conjectured to occur 
for the Gaussian theory. This result was suggested in~\cite{East:2021bqk}, 
but we strengthen it here by proving that the breakdown in hyperbolicity 
is gauge invariant and also by providing the complete picture of the phase 
space of end states.
These results indicate that the phase space available for spontaneous scalarization 
and descalarization, observed e.g.~in Refs.~\cite{Elley:2022ept,Silva:2020omi} without
back-reacting the scalar field onto the metric, might be quite small and require 
fine tuning of initial data.

When considering the perturbations of an (otherwise stable) boson star in its ground state by an 
infalling Gauss-Bonnet scalar field, we showed that the structure of the phase space is very similar to that of critical gravitational collapse with Gauss-Bonnet collapse initial conditions.
This result indicates that larger values of the coupling constant can push compact
objects, such as white dwarfs and neutron stars, into the strongly-coupled regime
(and hence out of its regime of predictability). Moreover, our results indicate that 
the formation of spontaneously scalarized black holes from the collapse of stars may require
fine-tuned initial data. 

When considering perturbations of an (otherwise stable) hair black hole, we were able to
study the size of the smallest possible 
BHs\footnote{The strength of the initial perturbation controls the 
breakdown of hyperbolicity therefore the dynamical stability of 
compact objects is tied with the kind of initial perturbation one considers.} 
allowed in both the shift symmetric and the Gaussian theory.
In the shift symmetric theory, we showed that the smallest possible BHs lie 
above the existence line obtained by static analysis.
For the Gaussian theory, we showed that the size of the smallest possible BHs 
lie very close to the Schwarzschild instability line, again confirming that 
spontaneous scalarization might require fine tuning of initial data 
and coupling constants~\cite{Kuan_2021}.
Our results indicate that the kind of initial data considered is very 
important when studying gradient expansions of gravity, such as in ESGB theory.

Our work points to a few natural avenues to explore in the future.
One direction would be to use our gauge invariant formalism to study the formation of 
NER in theories that include terms such as 
$\alpha(\phi)\left(\left(\nabla \phi\right)^2\right)^2$ in the action~\cite{Figueras:2020dzx,CCZ4-2022}, which
we set to zero in this paper.
It would be very straightforward to extend our derivation to include these 
terms. Indeed, recent work has investigated in detail the hyperbolic properties
of ``$K$-essence'' theories, which include terms like $\alpha\left(\phi\right)$
\cite{Bezares:2020wkn,Lara:2021piy}.


A second direction would be to extend the analysis performed here to 
axisymmetric gravitational collapse and obtain sufficient
conditions for the breakdown of hyperbolicity.
Our gauge invariant approach made heavy use of spherical symmetry to trade second derivatives 
of the metric for second derivatives of the scalar field. 
This is justified in spherical symmetry because the gravitational degrees of 
freedom are gauge degrees of freedom. 
It is not entirely clear as to how one could extend our 
approach to axisymmetry since  
propagating gravitational degree of freedom mix with scalar degrees of freedom. %
 
Spinning black holes are smaller than non-spinning black holes with the
same mass. 
Therefore, the addition of angular momentum may 
further shrink the size of 
parameter space available for stable evolution of BHs in ESGB gravity.
Studying the impact of rotation 
would also allow for the exploration of
ESGB gravity theories that allow for the spin-induced spontaneous scalarization of black holes~\cite{Elley:2022ept,Dima:2020yac,Doneva:2020nbb,Herdeiro:2020wei,Berti_2021}.
It would be interesting to explore if this phenomena also falls in the region 
of the parameter space where the theory loses hyperbolicity fails.

Finally, we emphasize that a more general analysis of the characteristic
polynomial of ESGB gravity (among other theories)
was carried out in \cite{Reall:2021voz}. It would be interesting
to compare our results in detail to those obtained in that work,
which may provide guidance on how to address some of the projects
we mentioned above.
\begin{acknowledgements}
We thank Aron Kovács and Helvi Witek for reading a preliminary version of the draft
and for providing insightful comments.
We are grateful to Jorge Noronha, Frans Pretorius, Harvey Reall and Leo Stein for reading a preliminary version of the draft and for providing valuable feedback.
AH, JR and NY acknowledge support from the Simons Foundation through Award number 896696.

\end{acknowledgements}
\appendix
\section{Spherical decomposition of the metric and construction of $P_{ab}$}
\subsection{Metric, Christoffel symbol and components of curvature tensor}\label{appendix:curvature-components}

In this section we list the Christoffel symbols and the curvature components obtained 
from the metric decomposition~\eqref{eq:spherical-metric}.
The Christoffel symbols are given by~\cite{Maeda_2008}
\begin{align}
    \Gamma^{a}_{bc} &= {}^{(2)}\Gamma^{a}_{bc}\\
    \Gamma^{a}_{AB} &= -\Omega_{AB} r D^{a} r \\
    \Gamma^{A}_{Bc} &= \frac{\delta^{A}_{B}}{r} D_c r\\
    \Gamma^{A}_{BC} &= {}^{(s)}\Gamma^A_{BC}\\
    \Gamma^{a}_{Ac} &= \Gamma^{A}_{ac} = 0\,.
\end{align}
Curvature components are given by~\cite{Maeda_2008}
\begin{align}
    R_{abcd}
    &= \mathcal{R}_{abcd} = 
    \frac{1}{2}\mathcal{R}\left(\alpha_{ac}\alpha_{bd} - \alpha_{ad}\alpha_{bc}\right)
    ,\\
    R_{aAbB} 
    &= -\left(rD_a D_b r\right) \Omega_{AB}
    \,\\
    R_{ABCD} 
    &= 
    \left[1 - (Dr)^2\right]r^2\left(\Omega_{AC}\Omega_{BD} - \Omega_{AD}\Omega_{BC} \right)
    ,
\end{align}
where $\mathcal{R}_{abcd}$ and $\mathcal{R}$ denote the two dimensional curvature tensor 
associated with the metric $\alpha_{ab}$.
The components of the double dual of the Riemann tensor are given by
\begin{align}
    \left({}^{*}R^{*}\right)_{abcd} 
    &= 
    \frac{\left[(Dr)^2 -1 \right]}{r^2}\left(\alpha_{ac}\alpha_{bd} - \alpha_{ad}\alpha_{bc}\right)
    \\
    \left({}^{*}R^{*}\right)_{aAbB} 
    &= 
    r\, \Omega_{AB}\left(\alpha_{ab}D^2r - D_{a}D_{b}r \right)
    \\
    \left({}^{*}R^{*}\right)_{ABCD} 
    &= 
    -\frac{\mathcal{R} r^4}{2}\left(\Omega_{AC}\Omega_{BD} - \Omega_{AD}\Omega_{BC} \right)
    \,.
\end{align}
The components of contracted curvature scalars are given by~\cite{Maeda_2008}
\begin{align}   
    \label{eq:ricci_a_b}
    R_{ab} &= \frac{1}{2}\mathcal{R}\alpha_{ab} -\frac{2}{r}D_a D_b r\\
    R_{AB} &= \left[1- (Dr)^2 - rD^2r\right]\Omega_{AB}\\
    G_{ab} &= \alpha_{ab} \left[\frac{2}{r}D^2r + \frac{(Dr)^2 -1}{r^2} \right] - \frac{2}{r}D_{a}D_{b}r \\
    G_{AB} &= \Omega_{AB}\left[rD^2r - \frac{r^2}{2}\mathcal{R} \right]\\
    R &= \mathcal{R} + \frac{2}{r^2}\left[1 - (Dr)^2\right] -\frac{4}{r}D^2r\,\\
    \label{eq:GB-decomp}
    \mathcal{G} &=\frac{4}{r^2}\left[ 2\left(D^2r\right)^2 - 2\left(D_aD_br\right)^2 + (1 - (Dr)^2)\mathcal{R}\right]\,,\\
    &= \frac{8}{r^2}\, D^a \left[D_a r D^2r - \left(D_b r\right) D_a D^b r\right] + \frac{\R}{r^2} \,.
\end{align}
The Gauss-Bonnet invariant is topological (a total derivative)
as we can see from the above expression (as the 2-D Ricci scalar is topological--note
from the metric determinant $\sqrt{-g}= r^2\sqrt{\alpha}\sqrt{\Omega}$).
We also note that the Ricci scalar $R_{ab}$ and the Gauss-Bonnet scalar $\mathcal{G}$ can be written down in following equivalent form using the STF operators defined in Eq.~\eqref{eq:def-lambda-stf}
\begin{align}
    \label{eq:Ricci-simplified}
    R_{ab} &= -\frac{2}{r} \lambda_{<ab>}[r] + \frac{\alpha_{ab}}{2}\left(\R - \frac{2}{r}\lambda[r]\right)\,, \\
    \label{eq:GB-simplified}
    \mathcal{G} &= \frac{4}{r^2} \lambda[r]^2 - \frac{8}{r^2}\lambda_{<ab>}{[r]}\lambda^{<ab>}[r] -\frac{4\sigma}{r^2}\mathcal{R}
    ,
\end{align}
where the scalar function $\sigma$ is defined in Eq.~\eqref{eq:sigma-def}.
\subsection{Projection of gravitational equations of motion}\label{appedix:proj-eom}
We start by projecting Eq.~\eqref{eq:grav-equations} 
onto the indices $(a,b)$
\begin{align}
    E_{ab} &= E_{<ab>} + \frac{1}{2}\alpha_{ab} E_2 \,,
\end{align}
where
\begin{align}
    E_{<ab>} &= -\frac{2}{r} \lambda_{<ab>}[r] -T_{<ab>} \nonumber\\
    &+ \left[\frac{16 \ell^2 \left(D^cr D_cf \right)}{r^2} \lambda_{<ab>}[r] + \frac{8\ell^2\sigma}{r^2}  \lambda_{<ab>}[f]\right] \,,\\
    E_2 &= \frac{\lambda[r]}{r} + \frac{\sigma}{r^2} - \left[\frac{8 \ell^2 \left(D^cr D_cf \right)}{r^2} \lambda[r] + \frac{4\ell^2\sigma}{r^2}  \lambda[f]\right] \nonumber \\
    &- T_2\,.
\end{align}
These equations can be solved for the trace-free and trace parts to obtain
\begin{align}
    \label{eq:lambda-stf-sol}
    \lambda_{<ab>}[r] &= \frac{4\ell^2\sigma}{\mu}\lambda_{<ab>}[f] - \frac{r^2}{2\mu} T_{<ab>} \,,\\
    \label{eq:lambda-sol}
    \lambda[r] &= \frac{\left(4\ell^2 \lambda[f] -1\right) \sigma}{\mu} + \frac{r^2 T_2}{\mu}\,.
\end{align}
The projection of equations onto indices $(A,B)$ results in 
\begin{align}
    E_{AB} 
    &=  
    \Omega_{AB}\left\{
        -
        \frac{r}{2}\mathcal{R}\mu 
        + 
        r\lambda[r]  
        + 
        8\ell^2 r \lambda_{<cd>}[r]\lambda^{<cd>}[f] 
        \right. 
        \nonumber\\
    &\left.- 4\ell^2 r \lambda[r]\lambda[f]\right\} - T_{AB} = 0
    .
\end{align}
Contracting with $\Omega^{AB}$ and solving for $\mathcal{R}$ we obtain 
\begin{align}\label{eq:ricci-2-sol}
    \mathcal{R} 
    &= 
    \frac{16\ell^2}{\mu} \lambda_{<cd>}[f]\lambda^{<cd>}[r] 
    - 
    \frac{8\ell^2}{\mu}\lambda[f]\lambda[r] 
    + 
    \frac{2 \lambda[r]}{\mu} 
    \nonumber\\
    &-\frac{r\Tilde{T}}{\mu}\,.
\end{align}
We can now use Eqs.~\eqref{eq:lambda-stf-sol}, \eqref{eq:lambda-sol} \& \eqref{eq:ricci-2-sol} 
to write the Gauss-Bonnet scalar~\eqref{eq:GB-simplified} as
\begin{align}
    \label{eq:GB-reduced}
    \mathcal{G} 
    &= 
    \frac{12}{r^2} \lambda[r]^2 
    - 
    \frac{24}{r^2}\lambda_{<ab>}{[r]}\lambda^{<ab>}[r] 
    - 
    \frac{8}{\mu}\lambda^{<ab>}[r] T_{<ab>} 
    \nonumber \\
    &
    - 
    \frac{8 T_2}{\mu}\lambda[r] + \frac{4 \sigma}{\mu r^3} \Tilde{T}
    \,.
\end{align}
The above equation can be simplified into 
\begin{align}\label{eq:gb-func-f}
    \mathcal{G} &=\frac{192 \lambda[f]^2 \sigma ^2 \ell ^4}{\mu ^2 r^2}\,\nonumber\\
    &+\frac{32 \lambda[f] \sigma  \ell ^2 \left(2 r^2 T_2-3 \sigma \right)}{\mu ^2 r^2}\,\nonumber\\
    &-\frac{384 \lambda_{<ab>}[f]\lambda^{<ab>}[f] \sigma ^2 \ell ^4}{\mu ^2 r^2}\,\nonumber\\
    &+\frac{64 \lambda_{<ab>}[f]T^{<ab>} \sigma \ell ^2}{\mu ^2}\,\nonumber\\
    &-\frac{2 r^2 T_{<ab>}T^{<ab>}}{\mu ^2}\nonumber\\
    &-\frac{2 \left(-2 r^5 T_2^2+8 r^3 \sigma  T_2-6 r \sigma ^2-2 \mu  \sigma  \Tilde{T}\right)}{\mu ^2 r^3}\,,
\end{align}
using Eqs.~\eqref{eq:lambda-stf-sol} and \eqref{eq:lambda-sol}\,.
\subsection{Details of the construction of the principal symbol}\label{appendix:pab-construction}

The definition of the principal symbol is given in Eq.~\eqref{eq:principal-symobl-defintion}.
We note that because of spherical symmetry the scalar field dynamics and characteristics are effectively restricted to the ``$t-r$'' plane.
That is, we only consider characteristic covectors of the form
$\xi_{\mu}=\left(\xi_a,0\right)$ (the angular indexed components are zero). 
In spherical symmetry Eq.~\eqref{eq:principal-symobl-defintion} reduces to
\begin{equation}
    P(\xi) 
    = 
    P^{ab}\xi_{a}\xi_{b} 
    = 
    \frac{\partial E_{\phi}}{\partial \left(\partial_{a}\partial_{b} \phi \right)}\xi_{a}\xi_{b}
\end{equation}
where $E_{\phi}$ is defined in Eq.~\eqref{eq:E-phi-v1}.
Let us use the symbol $\mathcal{P}\left[X \right]$ 
to denote the principal part of a quantity $X$. 
As the scalar degree of freedom drives the evolution of ESGB gravity in spherical symmetry,
we consider $\mathcal{P}\left[E_{\phi} \right]$ as the candidate principal symbol for
those spacetimes.
From Eq.~\eqref{eq:E-phi-v1}
\begin{align}
     \mathcal{P}\left[E_{\phi}\right] &= \mathcal{P}\left[D^aD_a \phi\right] + \ell^2 f' \mathcal{P}\left[\mathcal{G}[f]\right]\nonumber \\
     \label{eq:ephi-p-1}
     &= \alpha^{ab}\xi_{a}\xi_{b} + \ell^2 f' \mathcal{P}\left[\mathcal{G}[f]\right]\,.
\end{align}
Let us now calculate $\mathcal{P}\left[\mathcal{G}[f]\right]$.
From Eqs.~\eqref{eq:lambda-stf-reduced} \& \eqref{eq:lambda-r-reduced} we see that
\begin{align}
    \label{eq:var-lambda-stf}
    \mathcal{P}\left[ \lambda_{<ab>}[r] \right] &= \frac{4 \ell^2 \sigma }{\mu} \mathcal{P}\left[ \lambda_{<ab>}[f] \right] = \frac{4\ell^2 \sigma f'}{\mu} X_{<ab>}\,,\\
    \label{eq:var-lambda-r}
    \mathcal{P}\left[ \lambda[r] \right] &= \frac{4\ell^2 \sigma}{\mu}\mathcal{P}\left[ \lambda[f] \right] = \frac{4 \ell^2 \sigma f'}{\mu} \xi^2_2 \,
\end{align}
where $X_{<ab>} = \xi_{<a}\xi_{b>}$ and $\xi_2^2 = \alpha^{ab} \xi_a \xi_b$.
Let us now simplify $\mathcal{P}\left[\mathcal{G}[f]\right]$~\eqref{eq:GB-v1} using the above equations
\begin{align}
    \mathcal{P}&\left[\mathcal{G}[f]\right] = \frac{24}{r^2} \lambda[r] \mathcal{P}\left[ \lambda[r] \right] - \frac{48}{r^2}\lambda_{<ab>}{[r]}\mathcal{P}\left[ \lambda^{<ab>}[r] \right] \nonumber \\
    & - \frac{8}{\mu}\mathcal{P}\left[ \lambda^{<ab>}[r] \right] T_{<ab>} -  \frac{8 T_2}{\mu} \mathcal{P}\left[ \lambda[r] \right] \nonumber\,, \\
    &= \frac{24}{r^2}\left(\frac{4\ell^2 \sigma f'}{\mu} \right) \lambda[r] \xi_2^2 - \frac{48}{r^2}\lambda^{<cd>}[r] \left(\frac{4\ell^2 \sigma f'}{\mu} \right) X_{<cd>} \nonumber\\
    &- \frac{8}{\mu}\left(\frac{4\ell^2 \sigma f'}{\mu} \right) T^{<ab>}X_{<ab>} - \frac{8}{\mu}\left(\frac{4\ell^2 \sigma f'}{\mu} \right) T_2 \xi_2^2 \nonumber\,,\\
    &= \frac{96 \ell^2 \sigma f'}{r^2\mu} \lambda[r] \xi_2^2 - \frac{192 \ell^2 \sigma f'}{r^2\mu} \lambda^{<cd>}[r] X_{<cd>} \nonumber \\
    \label{eq:pg-simplified}
    &- \frac{32 \ell^2 \sigma f'}{\mu^2}T^{<ab>}X_{<ab>} - \frac{32 \ell^2 \sigma f'}{\mu^2}T_2 \xi_2^2\,.
\end{align}
We can now use the above equation in Eq.~\eqref{eq:ephi-p-1} to get 
$P\left[E_{\phi}\right]\equiv P_{ab}\xi^a\xi^b$:
\begin{align}\label{eq:pab-v1}
    P_{ab} 
    &= 
    \alpha_{ab}\left[
        1 
        + 
        \frac{96 \ell^4 \sigma (f')^2}{r^2\mu} \lambda[r] 
        - 
        \frac{32 \ell^4 \sigma (f')^2}{\mu^2}T_2
    \right] 
    \nonumber \\
    &
    - 
    \frac{192 \ell^4 \sigma (f')^2}{r^2\mu} \lambda_{<ab>}[r] 
    - 
    \frac{32 \ell^4 \sigma (f')^2}{\mu^2}T_{<ab>}
    \,.
\end{align}
We can now trade the $\lambda_{<ab>}[r]$ term in the above equation for the 
\emph{four} dimensional Ricci tensor $R_{<ab>}$ using Eq.~\eqref{eq:Ricci-simplified}.
This finally simplifies $P_{ab}$ to 
\begin{align}
    P_{ab} 
    =& 
    \alpha_{ab}\left[
        1 
        + 
        \frac{96 \ell^4 \sigma (f')^2}{r^2\mu} \lambda[r] 
        - 
        \frac{32 \ell^4 \sigma (f')^2}{\mu^2}T_2
    \right] 
    \nonumber \\
    &+ \frac{96 \ell^4 \sigma (f')^2}{r\mu} R_{<ab>} - \frac{32 \ell^4 \sigma (f')^2}{\mu^2}T_{<ab>}\,\\
    \label{eq:pab-ricci-appendix}
    =& \alpha_{ab}\left[1 + \pi_1 \ell^4 \left(\lambda[r] - \frac{r^2 }{3\mu}T_2 \right)\right] \nonumber \\
    &+ \pi_{1} \ell^4 r \left( R_{<ab>} - \frac{r}{3\mu} T_{<ab>} \right)\,.
\end{align}
where $\pi_1$ is defined in Eq.~\eqref{eq:pi1-def}.
We also provide the following equivalent forms in terms of $\lambda[f]$ and $\lambda_{<ab>}[f]$ of the above equation using Eq.~\eqref{eq:lambda-stf-sol}-\eqref{eq:lambda-sol} which maybe be useful for numerical implementation
\begin{align}
    \label{eq:pab-ricci-lf-appendix}
    P_{ab} &= \alpha_{ab}\left\{1 + \pi_1 \ell^4 \left[ \frac{(4\ell^2 \lambda[f] - 1)\sigma}{\mu} + \frac{2r^2}{3\mu}T_2 \right] \right\} \nonumber \\
    &+ \pi_{1} \ell^4 r \left( R_{<ab>} - \frac{r}{3\mu} T_{<ab>} \right)\,\\
    &\Leftrightarrow \nonumber \\
    \label{eq:pab-ricci-lf-labf-appendix}
    P_{ab} &= \alpha_{ab}\left\{1 + \pi_1 \ell^4 \left[ \frac{(4\ell^2 \lambda[f] - 1)\sigma}{\mu} + \frac{2r^2}{3\mu}T_2 \right] \right\} \nonumber \\
    &+ \pi_{1} \ell^4 r \left( \frac{2r}{3\mu} T_{<ab>} - \frac{8
    \ell^2 \sigma}{r\mu} \lambda_{<ab>}[f] \right)\,.
\end{align}
\begin{figure*}[thp!]
    \centering
    \includegraphics[width = 1\columnwidth]{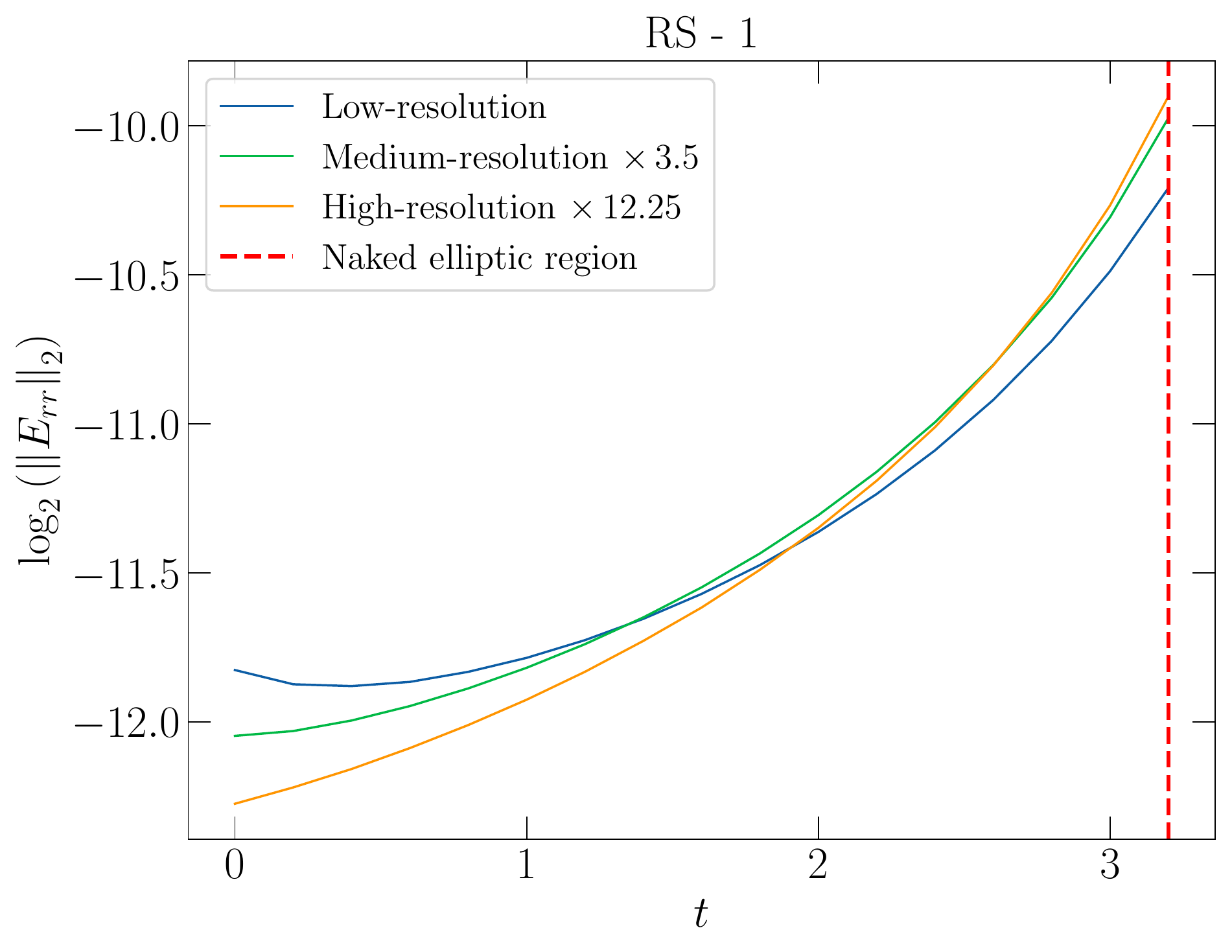}
    \includegraphics[width = 1\columnwidth]{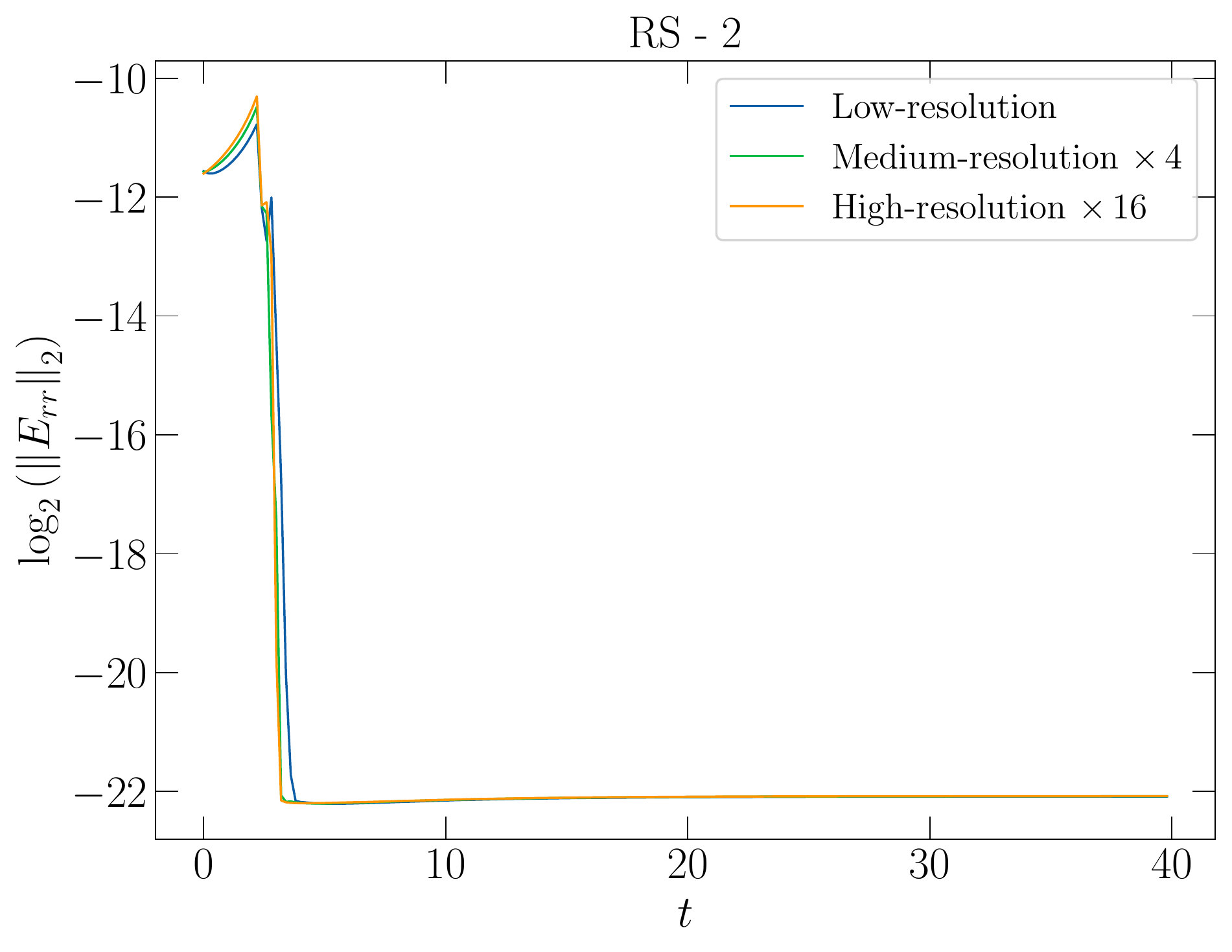}
    \includegraphics[width = 1\columnwidth]{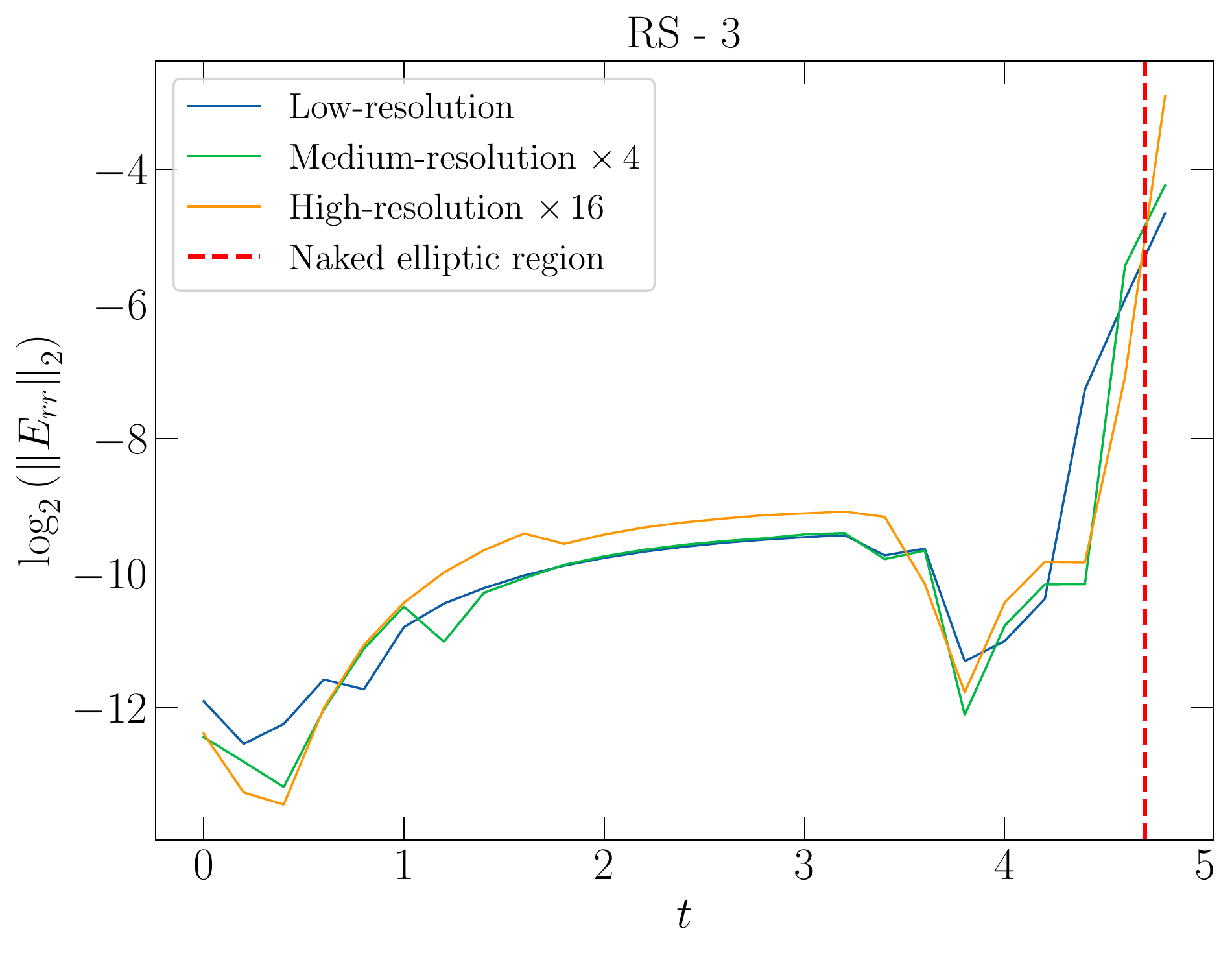}
    \includegraphics[width = 1\columnwidth]{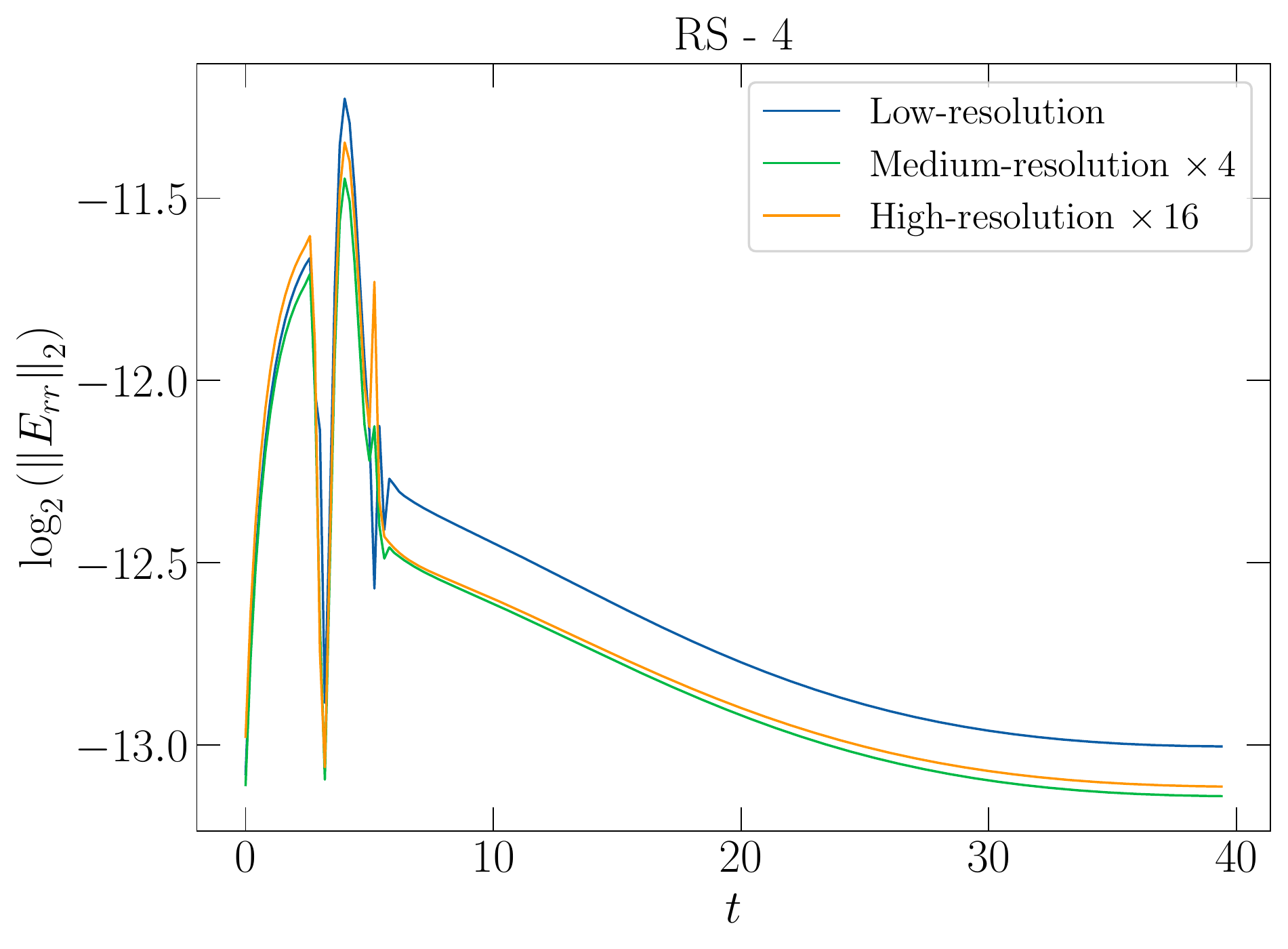}
    \caption{Convergence of the $rr$ component of the tensor equations of motion
    $E_{rr}$ (see Eq.~\eqref{eq:grav-equations}). 
    We achieve close to second order convergence for all our runs.}
    \label{fig:convg}
\end{figure*}
\section{Boson Star Solutions}\label{appendix:Boson-star}
Here we briefly review boson star solutions for the theory
\eqref{eq:boson_star_action} 
and summarize how we construct the boson star initial data.
For SBS initial data we are interested in how the boson star is affected by the presence of 
the Gauss-Bonnet scalar field. Therefore, we construct the boson star initial data in GR and 
then superimpose the Gauss-Bonnet scalar field later.
We will also only consider the boson star in it's ground 
state~\cite{Hawley_2000,Liebling:2012fv}.
Our coordinates are in Painlev\'{e}-Gullstrand coordinates
\begin{equation}
     ds^2 = -\alpha^2(1-\zeta^2)\, dt^2 + 2\,\alpha\, \zeta\, dt dr + dr^2 + r^2d\Omega^2\,.
\end{equation}
To construct the boson star initial data we use the following ansatz for the complex 
scalar field $\rho$
\begin{equation}\label{eq:boson-star-ansatz}
    \rho(t,r) = \rho_0(r) \exp\left(-i\omega(t + \upsilon(r))\right)
\end{equation}
where, $\upsilon(r)$ satisfies the following ODE
\begin{equation}
    \upsilon'(r) = -\frac{\zeta}{\alpha(1-\zeta^2)}\,,
\end{equation}
where the prime $'$ indicates a radial derivative:
$\upsilon':= d\upsilon/dr$.
We have defined $\sigma$ so that the boson star solution ansatz is computed in
Schwarzschild-type coordinates (the $\upsilon'$ essentially cancels out the 
rescaled shift variable $\zeta$ in the metric).
Note that we can think of the variable $\rho$ as giving us the ``density''
of the complex scalar field.
We next introduce dimensionless variables (we have defined
$Q_{(\rho)}:=d\rho/dr$)
\begin{align}
     \bar{r} &:= m_b r \,,\\
    \bar{\upsilon} &:= \omega \upsilon \,, \\
    \bar{\alpha} &:= \frac{m_b \alpha }{\omega} \,, \\
    \bar{Q}_{(\rho)} &:=  \frac{Q_{(\rho)}}{m_b}\,.
\end{align}
In these variables the field equations in GR reduce to 
\begin{align}
    \bar{\upsilon}' 
    &= 
    -\frac{\zeta}{\bar{\alpha}(1-\zeta^2)} 
    \\
    \zeta' 
    &= 
    -
    \frac{\bar{r} \bar{Q}_{(\rho)}^2}{4 \zeta }
    +
    \frac{\zeta }{2 \bar{r}}
    +
    \frac{1}{4} \bar{r} \bar{Q}_{(\rho)}^2 \zeta 
    -
    \frac{\bar{r} \rho _0^2}{4 \zeta }
    +
    \frac{\bar{r} \rho _0^2}{4 \bar{\alpha}^2 \zeta  \left(-1+\zeta ^2\right)}
    \\
    \bar{\alpha}' 
    &=
    \frac{\bar{r} \left(\bar{\alpha}^2 \bar{Q}_{(\rho)}^2 
    \left(-1+\zeta ^2\right)^2+\rho _0^2\right)}{2 \bar{\alpha} \left(-1+\zeta ^2\right)^2}
    \\
    \bar{Q}_{(\rho)}' 
    &= 
    -
    \frac{\frac{2 \rho _0}{\bar{\alpha}^2}
    +
    \frac{\left(-1+\zeta ^2\right) \left(2 \bar{r} \rho _0
    +
    \bar{Q}_{(\rho)} \left(-4+2 \zeta ^2+\bar{r}^2 \rho _0^2\right)\right)}{\bar{r}}}{2 \left(-1+\zeta ^2\right)^2} 
    \\
    \rho_0' &= \bar{Q}_{(\rho)}
    \,.
\end{align}
The above system of equations can be solved as a shooting problem with given $\rho_c$ 
and asymptotically flat boundary conditions
($\rho=0$, $\alpha=1$, and $\zeta=0$ at $r=\infty$).
The only ``free" parameter for the shooting method for the scaled variables is $\bar{\alpha}_0$. 
We therefore perform a search in $\bar{\alpha}_0$. 

The solution spectrum of boson star solutions are characterized by the number of 
nodes (zero crossings) in the profile of $\rho_0(r)$.
The ground state consists of no nodes and excited states consists of one or more nodes.
We will only consider boson stars in their ground state.
After the search for the ground state is finished for a given $\rho_c$, 
the value of the frequency is obtained by
\begin{equation}
    \frac{\omega}{m_b} = \frac{1}{\bar{\alpha}(\infty)}\,.
\end{equation}
After the frequency of the boson star is obtained we transform the variables back to 
scaled variables and obtain $\rho(0,r)$ from Eq.~\eqref{eq:boson-star-ansatz}. 
We use as this profile as the initial data for SBS.
We also use the following definition for the radius and the mass of the boson star
\begin{align}
    R_{\text{star}} = R_{95}\,,\\
    M_{\text{star}} = \left.\zeta^2 r/2 \right|_{r = \infty}\,,
\end{align}
where $R_{95}$ is the radius at which the density is $0.05$ times $\rho_c$. 
\section{Convergence tests}\label{appendix:Convg}
In this appendix we describe our code in more detail and the describe the convergence
of our simulations.
In the code, we compactify the radial coordinate with the following function~\cite{Ripley_2020_shift_symm_PG}
\begin{equation}
    r = \frac{ x}{\left(1 - \frac{x^2}{x_{\infty}^2} \right)}\,,
\end{equation}
where $x_{\infty}$ is the compactification length. 
We note that this form of compactification preserves the symmetry properties of the functions near $r=0$.
For all our simulations we set $x_{\infty} = 100$.
After compactification, we view the field variables $\left(\zeta,\alpha,P,Q,\phi\right)$ 
as functions of coordinates $(t,x)$.
We use a uniform grid in $(t,x)$ coordinates with a Courant–Friedrichs–Lewy (CFL) 
number of 0.2.
We discretize the spatial derivatives using second order finite difference stencils.
At the origin we stagger the grid and reflect the value of the function using the 
symmetry properties of the function~\eqref{eq:bcs-alpha-fs}-\eqref{eq:bcs-phi-fs}.
For BH spacetimes we discretize the spatial derivatives using forward stencils at 
the excision point up until 3 grid points before the location of the apparent horizon 
and use central stencils thereafter. 
We find that this strategy of using forward finite difference stencils reduces the 
oscillations one would observe when the elliptic region begins to grow for BHs near 
the threshold between evolution to stable scalarized BHs and naked elliptic regions. 

Our method for evolution for a single time step is as follows.
We first solve the constraint equations~\eqref{eq:constraint-M}-\eqref{eq:constraint-alpha} 
using Heun's method to obtain $\zeta$ and $\alpha$. 
After the integration of the constraint, we evolve the time evolution equations~\eqref{eq:evoution-P}-\eqref{eq:evolution-phi}. 
Our time stepping method uses 2nd order stencils for spatial derivatives 
followed by a RK4 time step of the discretized set of ODEs.
We continue the evolution until the system settles to a static state or 
we form a naked elliptic region.

We now present convergence results from 4 different runs in the shift symmetric theory
(see Fig.~\ref{fig:convg})
\begin{itemize}
    \item \textbf{RS-1}: 
    Run with CIC and parameters, $\ell = 0.5$ and $A = 0.1$ 
    which leads to the formation of a naked elliptic region.
    \item \textbf{RS-2}: Run with CIC and parameters, $\ell = 0.5$ and $A = 0.18$ 
    which leads to the formation of scalarized BH.
    \item \textbf{RS-3}: Run with BHIC and parameters $\ell = 0.5$ and $M = 1.1$ which leads to the formation of naked elliptic region outside the AH.
    \item \textbf{RS-4}: Run with BHIC and parameters, $\ell = 0.5$ and $M = 1.3$ which leads to the formation of a stable scalarized BH.
\end{itemize}

These four runs illustrate the possible end states of
gravitational collapse apart from dispersion back to flat space.
For each of these runs we use 3 different resolutions. 
The lowest resolution run has $n_x = 5000$ ($dx \sim 0.02$) points, 
and the the medium resolution run and high resolution run 
have double and quadruple number of radial points of the lowest resolution run.  
We use the $E_{rr}$ component of the gravitational equation of motion (see Eq.~\eqref{eq:grav-equations})as a measure of the rate of convergence.
From Fig.~\ref{fig:convg} we see that we achieve second order convergence 
for all our runs except for RS-1,
where we see slightly less than second order convergence. 
From our discretization scheme we expect an order of convergence between second and
four order (depending on what terms in our code contribute the most to our error budget).
We have checked that we achieve similar results for the Gaussian theory and for SBSIC intial data.
\bibliographystyle{apsrev4-1}
\bibliography{ref}
\end{document}